\documentclass[aps,prx,twocolumn,footinbib,floatfix]{revtex4-2}

\usepackage{CJK}
\usepackage{amsmath}
\usepackage{amsfonts}
\usepackage{bm}
\usepackage{soul}
\usepackage[caption=false,subrefformat=parens,labelformat=parens]{subfig}
\usepackage{color} 
\usepackage[table,dvipsnames]{xcolor}
\usepackage[colorlinks,linktocpage,bookmarks=false]{hyperref}
\usepackage{graphicx}%
\usepackage{dsfont}
\usepackage{booktabs}

\newcommand\myshade{85}
\definecolor{myrulecolor}{RGB}{150,20,0}
\colorlet{mylinkcolor}{violet}
\colorlet{mycitecolor}{YellowOrange}
\colorlet{myurlcolor}{Aquamarine}
\hypersetup{
	linkcolor  = myurlcolor!\myshade!black,
	citecolor  = mycitecolor!\myshade!black,
	urlcolor   =  myrulecolor!\myshade!black,
}

\usepackage{graphicx}
\graphicspath{{FIG/}}
\usepackage{multirow}
\usepackage{braket}

\newcommand{\beq}{\begin{equation}}
\newcommand{\eeq}{\end{equation}}
\newcommand{\bea}{\begin{eqnarray}}
\newcommand{\eea}{\end{eqnarray}}

\newcommand{\Tr}{\text{Tr}}

\newcommand{\bfq}{\bm{q}}
\newcommand{\bfr}{\bm{r}}
\newcommand{\bfu}{\bm{u}}

\newcommand{\bfnabla}{\bm{\nabla}} 

\renewcommand\[{\begin{equation}}
\renewcommand\]{\end{equation}}

\newcommand{\MSC}{${\mathrm{McSc}}_{2}{\mathrm{S}}_{4}$}

\makeindex

\begin{document} 
	\begin{CJK*}{UTF8}{gbsn} 
		\title{Low energy structure of spiral spin liquids}
		\author{Han Yan (闫寒)}
		\email{hy41@rice.edu}
		\affiliation{Rice Academy of Fellows } \affiliation{Department of Physics \& Astronomy, Rice University, Houston, TX 77005, USA}
		\author{Johannes  Reuther}
    \affiliation{Helmholtz-Zentrum Berlin f\"{u}r Materialien und Energie, Hahn-Meitner Platz 1, 14109 Berlin, Germany}
    \affiliation{Dahlem Center for Complex Quantum Systems and Fachbereich Physik, Freie Universit\"at Berlin, 14195 Berlin, Germany}
		\date{\today}
\begin{abstract}
In this work we identify a previously unexplored type of topological defect in spiral spin liquids -- the momentum vortex -- and reveal its dominant role in shaping the low energy physics of such systems.
Spiral spin liquids are a class of classical spin liquids featuring sub-extensively degenerate ground states.  
They are distinct from spin liquids on geometrically frustrated lattices, in which the ground state degeneracy is extensive and connected by local spin flips.
Despite a handful of  experimental realizations and many theoretical studies,
a concrete physical picture of their spin liquidity has not been established so far.
In this work, we study a 2D spiral spin liquid model to answer this question.
We find that the local momentum vector field can carry topological defects in the form of vortices, which, however, 
have very different properties from the commonly known spin vortices.
The fluctuations of such vortices lead the system into a liquid phase at intermediate temperatures.
Furthermore, the
effective low energy theory of such vortices indicates their
equivalence to quadrupoles of fractons in a rank-2 U(1) gauge theory or, alternatively, to quadrupoles of disclinations in elasticity theory.
At very low temperatures, 
the system freezes into a glassy state in which these vortices form a rigid network 
with straight-line domain walls.
Our work sheds light on the nature of spiral spin liquids 
and also paves the way toward understanding their quantum limit.
\end{abstract}
\maketitle
\end{CJK*}

\tableofcontents  
\section{Introduction}

A fascinating theme of condensed matter is emergence \cite{anderson72} -- from the Anderson--Higgs mechanism in superconductors, to topological defects in superfluid $^3$He, to gauge theories in the description of frustrated magnets -- each condensed matter system  is itself a different universe in a grain of sand.
In particular, 
spin liquids, both classical and quantum, have been proven to be a fruitful field to search for exotic phases, (classical analogs of) fractionalized excitations and topological orders  \cite{Anderson1973,Bramwell2001Science,hermele04PRB,Morris2009,Fennell2009,Morris2009,Balents2010,Savary2016,Benton2016,Zhou2017,Yan2017PhysRevB}.

A peculiar class of spin liquids is the spiral spin liquid which features sub-extensively degenerate classical ground states \cite{rastelli79PhysicaB,Chandra1988PhysRevB,Fouet2001}. 
Each ground state is a  spin spiral state with a certain wave-vector,
and these ground state wave-vectors form a ring (or other manifolds of lower dimension) in the reciprocal lattice,
instead of just isolated points in conventional magnetic systems. 
They have a long history of theoretical study  \cite{rastelli79PhysicaB,Chandra1988PhysRevB,Fouet2001,Seabra2008PRB,Mulder2010PhysRevB,Holt2014PhysRevB,Attig2017PhysRevB,Buessen2018PhysRevLett,Okumura2010JPSJ,Okubo2012PhysRevLett,Shimokawa2019PhysRevB,Shimokawa2019PhysRevLett,Yao2021Frontier,Niggemann2019JPCM,huang2021arXiv}, and have already been observed in a handful of materials, such as 
\MSC\ \cite{Gao2016NatPhys,Bergman2007NatPhys,Iqbal2018PRB},
${\mathrm{MgCr}}_{2}{\mathrm{O}}_{4}$ \cite{Bai2019PRL},
${\mathrm{CoAl}}_{2}{\mathrm{O}}_{4}$
 \cite{Zaharko2011PhysRevB},
 ${\mathrm{NiRh}}_{2}{\mathrm{O}}_{4}$
 \cite{Chamorro2018PhysRevMaterials},
 and 
  ${\mathrm{Ca}}_{10}{\mathrm{Cr}}_{7}{\mathrm{O}}_{28} $ \cite{Balz2016NatPhys,Balz2017PhysRevB,Sonnenschein2019PhysRevB,Kshetrimayum2020AOP,Pohle2021PhysRevB,pohle2017arxiv}.

However, 
a clear physical picture of their spin liquidity has not been established to our knowledge.
This is in contrast to spin liquids from geometric frustration (e.g. in pyrochlore spin ice systems),
in which the ground state degeneracy is of local nature and, hence, the mechanism of fluctuations is much better understood:
The system can visit different classical ground states by flipping only a few spins which can be easily accomplished by thermal fluctuations.
Spiral spin liquids have a much smaller ground state degeneracy and, as a consequence, local manipulations are not sufficient to bring the system into different ground states.
It is already known that in the thermodynamic limit, the spin configurations occupy all the ground state wave-vectors.
Yet, since changing from one ground state spiral wave-vector to another is a global action on the spins, 
how exactly all the ground state wave-vectors can be populated remains unclear.

In this work,
we thoroughly investigate the low energy behavior of 2D spiral spin liquids with XY spins to understand their spin liquidity.
We find that, besides the commonly known spin vortices,
there exists another type of topological defect that corresponds to
vortices in the local momentum vector field on the coarse-grained lattice.
These topological defects, dubbed \textit{local momentum vortices}, have very different mathematical properties compared to spin vortices 
and play a crucial role in determining the low energy behavior of spiral spin liquids. 

At intermediate temperatures,  
our effective continuum theory can be formulated in terms of the local momentum vector field,
and is found to be similar to that of elasticity and  scalar-charged rank-2 U(1) gauge theory \cite{XuPRB06,PretkoPRB16,Rasmussen2016arXiv,PretkoPRB17,PretkoPRL18,pretko19,Gromov2019PhysRevLett,Nguyen2020SciPost}.
The local momentum vector field plays the role of lattice distortions in elasticity, and  
their vortices are identified as vacancies/interstitials in elasticity or quadrupoles of fractons \cite{Nandkishore2019Review,Pretko2020Review}.
The liquid phase can then be understood as a state with mobile topological defects populating the system. 

Proceeding to lower temperatures,   
the defects continuously lose their mobility,
leading to a glassy state which we call a \textit{rigid network of momentum vortices}. 
In this network state, the momentum vortices sit at the vertices, and the edges connecting them are narrow domain walls between regimes of different momenta.
Particularly, due to the unusual low energy properties of momentum vortices, the domain walls must be straight lines, establishing the rigidity of the network.
All these phenomena are thoroughly analysed and numerically demonstrated.

Our study of spiral spin liquids answers the key  question of their low energy structure
and reveals a previously unexplored type of topological defect
with unexpected connections to fracton physics.
It also paves the way for understanding the quantum limit  of these models \cite{Fouet2001,Gong2013PhysRevB,Niggemann2019JPCM}, which may give rise to quantum spin liquids.
A particularly promising system to realize the phases studied here, but under the additional effects of quantum fluctuations, is the bilayer kagome material ${\mathrm{Ca}}_{10}{\mathrm{Cr}}_{7}{\mathrm{O}}_{28} $ \cite{Balz2016NatPhys} which has recently been identified as a quantum spin liquid candidate. 

\section{The spiral spin liquid model and its phase diagram}
In this work,
our main focus is on 2D spiral spin liquids
whose $T=0$ ground state degeneracy is homotopic to a ring. 
Our showcase example is the classical square lattice XY spin model with couplings up to third nearest neighbours.

However, this is not the only model exhibiting spiral spin liquid physics. 
There are other models defined on the honeycomb and triangular lattices with further neighbour couplings~ \cite{Fouet2001,Seabra2008PRB,rastelli79PhysicaB,Niggemann2019JPCM} in 2D, and also on the diamond~\cite{Bergman2007NatPhys}, face centered cubic~\cite{Attig2017PhysRevB,penc19,balla20}, and body centered cubic~\cite{Attig2017PhysRevB} lattices in 3D. A more general overview on the construction of these models can be found in Refs.~\onlinecite{Niggemann2019JPCM,Yao2021Frontier}.

\subsection{The square lattice XY spin model}
\begin{figure*}[t]
\centering
\includegraphics[width=0.99\linewidth]{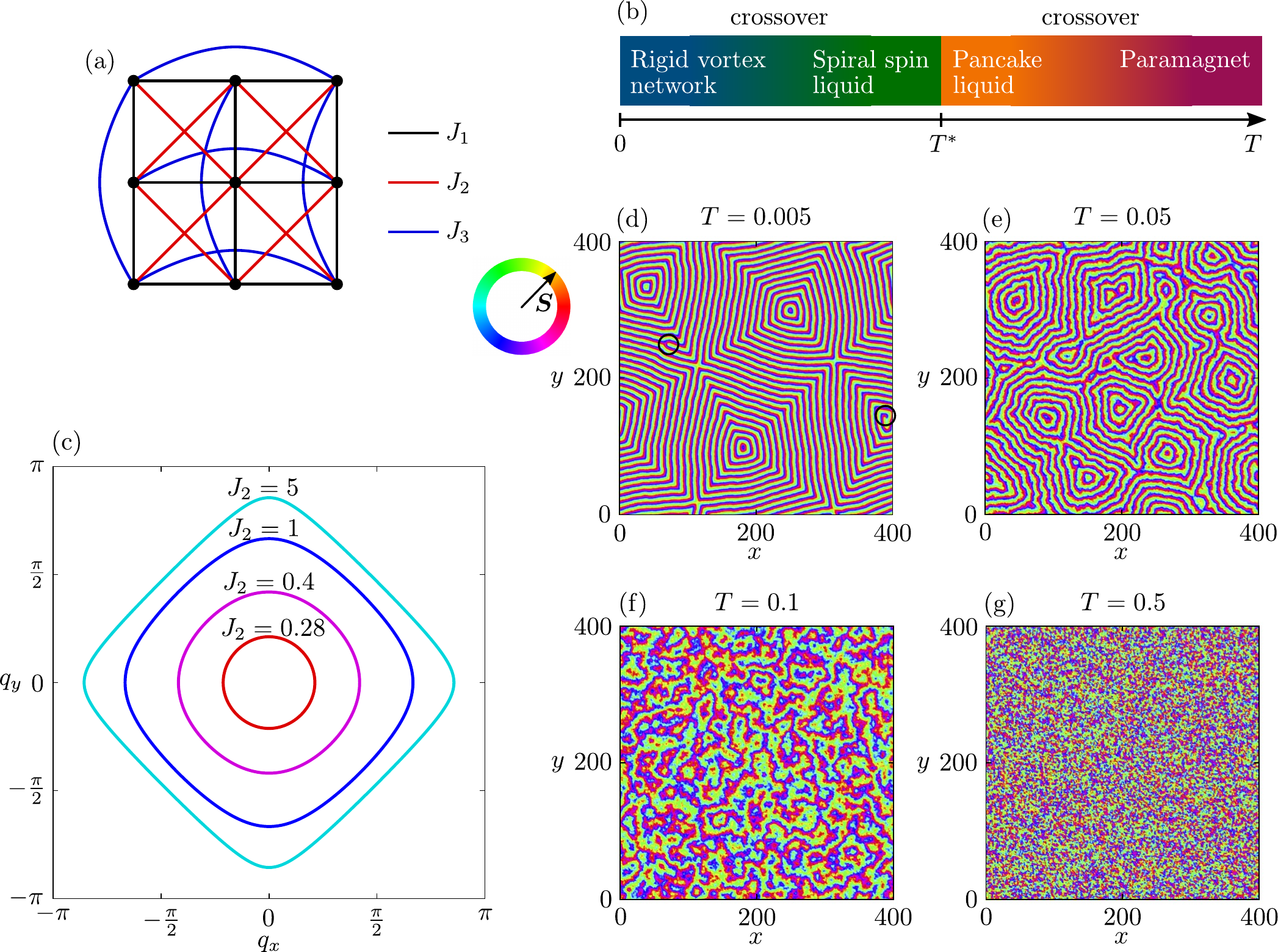}
\caption{(a) Definition of the couplings $J_1$, $J_2$, and $J_3$ on the square lattice. (b) Schematic phase diagram of the square lattice XY spin model as a function of temperature. (c) Shape of the spiral contour in momentum space [i.e. solution of Eq.~(\ref{EQN_GS_q_condition})] for varying $J_2$ and $J_3=J_2/2$. (d)-(g) Examples for spin configurations in the four temperature regimes of the square lattice XY model: (d) rigid vortex network, (e) spiral spin liquid, (f) pancake liquid, (g) paramagnet. The in-plane direction of the spins is color-encoded. The spin states are obtained in Monte-Carlo simulations of a $400\times400$ square lattice XY model with $\delta=0.03$. The black circles in (d) mark the positions of spin vortices.} 
\label{fig1}
\end{figure*}

The Hamiltonian for the square lattice XY spin model is given by
\begin{equation}
\label{EQN_Ham_SQXY}
{\cal H}_\text{sq-XY}
= J_1 \sum_{\langle ij \rangle_1} \ {\bm S}_i \cdot {\bm S}_j 
+ J_2 \sum_{\langle ij \rangle_2}  {\bm S}_i \cdot {\bm S}_j
+ J_3 \sum_{\langle ij \rangle_3}  {\bm S}_i \cdot {\bm S}_j.
\end{equation}
Here, ${\bm S_i}$ are normalized ($|{\bm S}_i|=1$) classical XY (i.e. two-component) spins and $J_{1,2,3}$ are first, second, and third nearest neighbour couplings  respectively, see Fig.~\ref{fig1}(a).
In the region of parameters
\[
J_1 = -1,\ J_3 = J_2/2, \ J_2 > 1/4,
\label{EQN_parameter_region}
\]
the ground states are spin spirals of momentum $\bfq$,
\begin{align}\label{spin_spiral}
{\bm S}_i&=(\cos(\Phi({\bm r}_i)),\sin(\Phi({\bm r}_i))\notag\\
&=(\cos({\bm q}\cdot{\bm r}_i+\phi),\sin({\bm q}\cdot{\bm r}_i+\phi))
\end{align} 
where $\bfr_i$ is the position of site $i$, and $\phi$ corresponds to a global rotation of all spins \cite{rastelli79PhysicaB}. Most importantly, the system exhibits a degenerate set of ground state spirals with momenta $\bfq$ satisfying the condition
\[  
\label{EQN_GS_q_condition}
2\cos^2(q_x) 
+  2\cos^2(q_y)+4\cos(q_x)\cos(q_y)   
= \frac{1}{2 J_2^2}.
\]
The solutions $\bfq$ form a continuous 1D manifold isomorphic to a loop around the Brillouin zone center. The shape of the manifold depends on the value of $J_2$, and is illustrated in Fig.~\ref{fig1}(c). In the limit of small 
\[
\delta\equiv J_2-\frac{1}{4}\ll1
\]
the spiral contour shrinks and becomes circular, obeying  $q=|{\bm q}|=4\sqrt{\delta}$, until for $J_2 < 1/4$ a simple ferromagnet is realized. 

An important property of XY models with this type of ground state degeneracy is that they exhibit two distinct types of U(1) symmetries. The first one is the standard global U(1) spin rotational symmetry which is generated by a simultaneous rotation of all spins, $\Phi({\bm r}_i)\rightarrow \Phi({\bm r_i})+\alpha$. The second one is a U(1) symmetry in momentum space which changes the momentum ${\bm q}$ of a ground state spiral along the contour of solutions of Eq.~(\ref{EQN_GS_q_condition}). Note that this second symmetry is in principle only a property within the exact ground state manifold. However, in the limit of small $\delta$ {\it and} small momenta ${\bm q}$ where the system is approximately spherical, the effective low-energy U(1) symmetry still stands.

The consequences of U(1) symmetry in spin space are well understood. 
It can trigger a finite temperature Kosterlitz-Thouless transition associated with a proliferation of spin vortices~\cite{Kosterlitz,KosterlitzThouless}. 
On the other hand, the effective momentum U(1) symmetry is much less studied. 
While previous works, mostly focusing on numerics, have found that it gives rise to a spiral spin liquid phase where the system thermally fluctuates through the degenerate manifold of spiral states, 
the precise mechanism behind such fluctuations is poorly understood. 

We emphasize that this  mechanism causing the spin liquid physics is very different from  other better-understood classical spin liquid models on frustrated lattices (e.g. pyrochlore spin ice \cite{hermele04PRB,Harris1997PhysRevLett}).
Those models are endowed with {\it local} zero modes giving rise to an extensive ground state degeneracy.
Hence, it is very intuitive that at small but finite temperatures,  
local spin flips enable the system to visit the degenerate manifold of ground states leading to a classical spin liquid phase.
The situation in spiral spin liquids is very different.
Even though the spiral ring degeneracy is sub-extensively large,
a change of momentum ${\bm q}$ is still a global operation associated with the modification of a macroscopic number of spins. 
It is, hence, a priori unclear how thermal fluctuations may induce a liquid-like property in the system studied here.

\subsection{Summary of main results}
Before we study spiral spin liquids in detail in the next two sections, we give a brief outline of our main conclusions. Although we concentrate here on the square lattice XY model only, we expect our results to also apply to other models with a spiral degeneracy.

\begin{figure}[t]
\centering
\includegraphics[width=0.99\columnwidth]{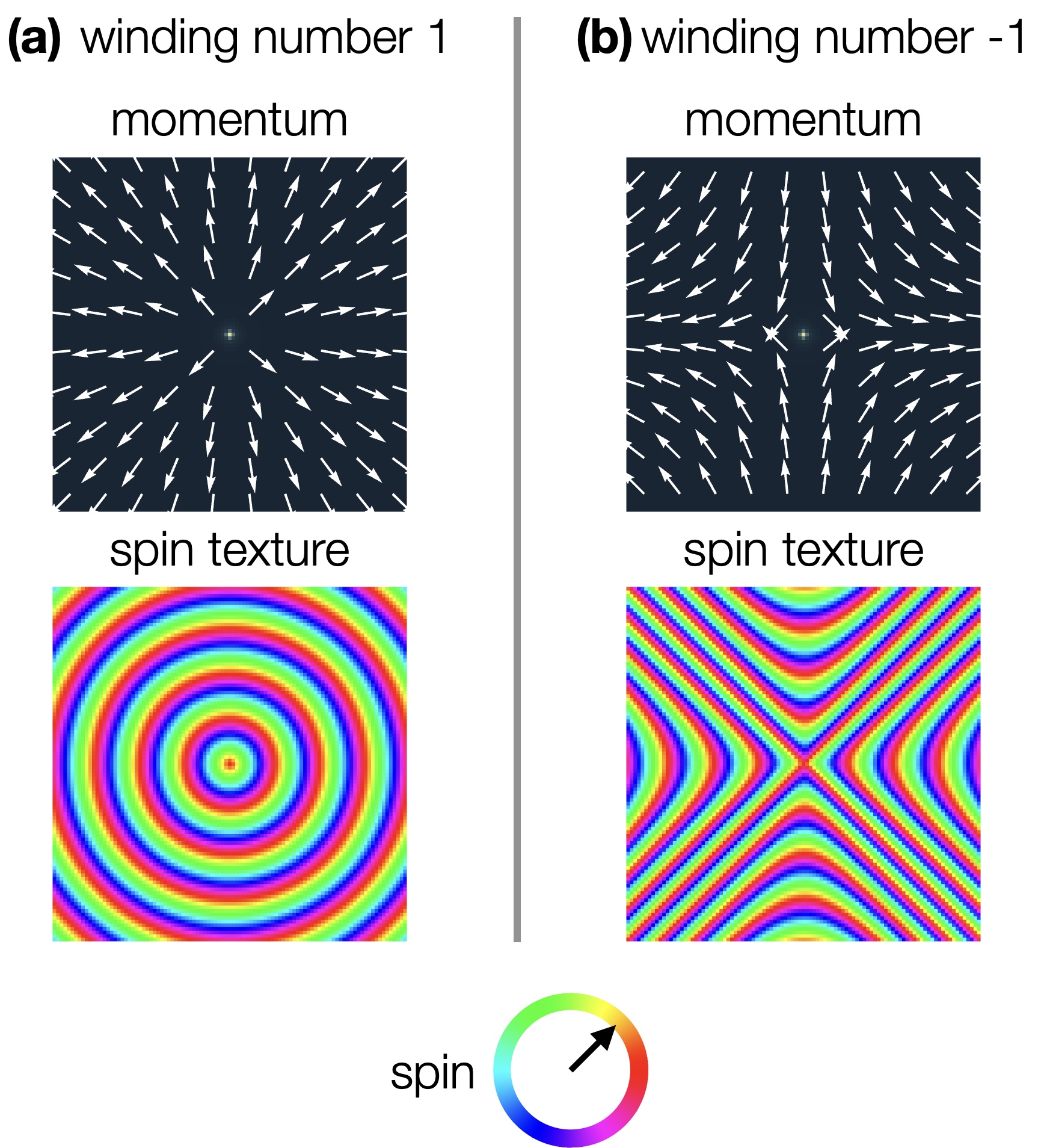}
\caption{Ideal examples of momentum (anti-) vortices with winding numbers $n=1$ (left column) and $n=-1$ (right column). The top panel shows  the momentum $\bfq = \bfnabla \Phi(\bfr)$ configuration, and the bottom panel shows the spin texture $\Phi(\bfr)$.
\label{fig:vortices_examples}}
\end{figure}

First, our core result is the identification of topological defects from the spiral ring U(1) symmetry (\textit{not} the spin rotation symmetry).
In these topological defects, the spin configuration $\Phi(\bm{r})$ (where $\Phi$ is the spins' inplane angle) varies smoothly without singularities. 
However, the corresponding coarse-grained momentum
\[
\bfq = \bfnabla \Phi(\bfr),
\]
which takes its value only in the neighborhood of the spiral ring [Fig.~\ref{fig1}(c)] and varies in space,
can have a non-trivial winding around a loop.
As one would expect, these momentum vortices (see Fig.~\ref{fig:vortices_examples} for two  vortex configuration examples)
play a central role in shaping the low temperature physics.
Here, we also note that the properties of such momentum vortices are very different from the commonly-known spin vortices, due to the fact that in the absence of spin vortices, the momentum field is subject to a curl-free condition. 
As we will elaborate later, a peculiar consequence of this restriction is that momentum vortices can only be realized with winding numbers $n\leq1$.

Our numerical Monte Carlo results for small $\delta$ indicate four different temperature regimes, see phase diagram in Fig.~\ref{fig1}(b) and exemplary spin configurations in Fig.~\ref{fig1}(d)-(g). At large temperatures a trivial paramagnet is realized where the spins can be considered as uncorrelated [Fig.~\ref{fig1}(g)]. Upon cooling, the system first undergoes a crossover into a phase referred to as `{\it pancake liquid}' in Ref.~\onlinecite{Shimokawa2019PhysRevLett}, see Fig.~\ref{fig1}(f). While this regime already shows a certain degree of correlations between spins, the thermal fluctuations are too strong to restrict the momentum to the spiral contour. As a result, spiral configurations do not form, as is indicated by the absence of clear stripe-like patterns in Fig.~\ref{fig1}(f).

When decreasing the temperature further, the system undergoes a transition at $T=T^*$ into a regime with well-defined spin spirals which can be recognized as stripy configurations in Figs.~\ref{fig1}(d) and ~\ref{fig1}(e). The investigation and characterization of the spiral regime at $T<T^*$ represents the main subject of our work. Despite the common spiral motif, the physical properties near the upper ($T\lesssim T^*$) and lower ($T\ll T^*$) boundaries are distinctly different. We refer to the two regimes at $T<T^*$ as `{\it spiral spin liquid}' (or alternatively as `{\it R2-U1/elasticity phase}') and `{\it rigid vortex network}' and discuss them separately in Secs.~\ref{SEC_r2u1} and \ref{SEC_rigid_fracton_network}, respectively.

In the spiral spin liquid phase [Figs.~\ref{fig1}(e)] the local momentum ${\bm q}$ is approximately confined along the spiral contour, however, the direction of ${\bm q}$ fluctuates strongly in real-space and Monte-Carlo time, establishing a liquid-like property. Most importantly, momentum (anti-) vortices with winding numbers $n=\pm1$ are clearly discernible and represent the key source of fluctuations in this regime (for comparison, see the ideal examples of momentum vortices with $n=\pm1$ in Fig.~\ref{fig:vortices_examples}). The occurrence of momentum vortices prompts us to analytically investigate their precise nature and contrast them with spin vortices. 

Another surprising property of the spiral spin liquid at $T\lesssim T^*$ is that its effective continuum theory for local momentum $\bfq$ can be mapped onto the elasticity theory
of the shear modulus. 
In turn, elasticity theory is known to be dual to a rank-2 U(1) fracton tensor gauge theory~\cite{PretkoPRL18,pretko19} in which immobile and scalar fractons (i.e., charges of the gauge theory) correspond to lattice disclinations and subdimensional fracton dipoles are related to lattice dislocations. This justifies the name R2-U1/elasticity phase. We show, however, that the nature of the momentum degree of freedom ${\bm q}$ in our spiral model and, particularly, the integer-quantized spin vortices do not allow for the existence of isolated fractons and dipoles of fractons. 
Instead,
the momentum vortices of winding number $n=1$ correspond to fracton {\it quadrupoles} in R2-U1 theory, or vacancies or interstitials in elasticity.
Hence, the spiral spin liquid phase can be effectively understood as a charge- and dipole-free rank-2 U(1) tensor gauge theory.
Similarly, momentum vortices of winding number $n<0$ can be understood as higher multipoles of vacancies and interstitials.  
The tensor gauge theory property of spiral spin liquids is also investigated numerically by demonstrating the occurrence of four-fold pinch points~\cite{PremPhysRevB.98.165140} in the electric-field correlator (Fig.~\ref{fig:mc_results1}).

The spiral spin liquid exhibits both  spatial fluctuations in the momentum direction ${\bm q}/q$ and in the momentum amplitude $q$ (where $q$ remains in the vicinity of the spiral contour, $q\approx4\sqrt{\delta}$). Since both variations in space  cost energy, they smoothly freeze out as one further decreases the temperature. 
In our numerical results [see Figs.~\ref{fig1}(d)] this freezing occurs via the formation of spiral domains each characterized by a well-defined momentum direction. In such states, the excitation energy from spatial variations of ${\bm q}$ is completely concentrated along narrow domain walls which form a rigid network spanning the entire system. 
Most importantly, momentum vortices correspond to the intersections of domains walls. 
The system also exhibits thermally excited spin vortices  which are marked by circles in Fig.~\ref{fig1}(d).
However, they are associated with much larger excitation energies than momentum vortices. 
No indications for a Kosterlitz-Thouless transition and a binding into low-temperature vortex-antivortex pairs is found, neither in spin- nor in momentum space. 
We explain this by the special restrictions imposed on the effective degrees of freedom such as the curl-free condition for ${\bm q}$.

Our numerical results at the lowest simulated temperatures also indicate that the domain wall network preferably forms rectangular patterns. 
This can  partially  be explained by the underlying square lattice nature of our model but is mainly due to the antivortices ($n=-1$) which realize the lowest excitation energy cost when four domain walls are radiating from the vortex core. 
Their characterization through the number and precise arrangement of radiating domain walls allows us to develop a general classification scheme for momentum vortices. 
Due to Mermin-Wagner theorem the formation of rectangular domains associated with the breaking of momentum symmetry [$U(1)\rightarrow\mathds{Z}_4$] does not occur in a finite-temperature transition but rather in a smooth crossover separating the spiral spin liquid and the rigid vortex network regimes. 
Particularly, with decreasing temperature the average domain size continuously increases while thermalization and mobility of momentum vortices significantly slow down. 
Besides the shear modulus term in the system's continuum theory, further contributions not allowed in elasticity theory (such as a potential term for the momentum) become increasingly important in this low-temperature regime. 
As a result, a rank-2 U(1) gauge theory description is no longer appropriate which is also seen in the fading of the four-fold pinch points in the electric field correlator (Fig.~\ref{fig:mc_results1}).

\section{The R2-U1/elasticity phase}\label{SEC_r2u1}

In this section, we discuss the properties of the spiral spin liquid as well as its relation to a rank-2 U(1) gauge theory and elasticity theory. Since fluctuations of momentum vortices are the main driving force behind this phase, we start discussing their precise nature and the constraints imposed by the curl-free condition. At the end of the section, we confirm our findings with numerical results. 
 
\subsection{Vortices of local momentum vector field}

Let us first qualitatively describe the arising of vortices from the local momentum vector field. We view the lattice in a coarse-grained way. Each coarse-grained block is over a few lattice sites so that we can define the local momentum vector ${\bm q}$ as the gradient of the spins' in-plane angle $\Phi$, 
\[
\label{EQN_q_vec_definition}
\bfq = \bm{\nabla} \Phi.
\]
At the same time the block is not too big so that the momentum vector effectively does not vary within the block.
In this way, we have defined a vector field $\bfq(\bfr)$ of the local momentum on the coarse-grained lattice.

At low temperatures, each momentum vector takes a value from the spiral momentum ring, or at least within a narrow region around it.
Consequently, the configuration space of a momentum vector at each individual coarse-grained block is homotopic to a circle,
so the winding number $n$ of the momentum vector field is a well-defined topological quantity on an arbitrary loop on the lattice.
Examples of winding number $\pm 1$ vortices are shown in Fig.~\ref{fig:vortices_examples}.

Momentum vortices are distinctly different from spin vortices which are defined in terms of spin winding along closed loops. In fact, spin and momentum vortices are independent of each other. Furthermore, unlike normalized spins,  there is no strict amplitude constraint on the momentum ${\bm q}$. 
As a consequence, momentum vortices can be realized on continuously varying spin textures without any singularity. 
Another drastic and, at a first glance, unexpected difference concerns the possible winding numbers $n$ of momentum vortices which we will elaborate on in the next subsection.

\subsection{Curl-free constraint on momentum vortices}
\label{SEC_winding_number_constraint}

In the absence of spin vortices,
$\Phi$ varies continuously on every point of the system.
So $\bfq(\bfr)$ must obey the curl-free condition
\[
\label{EQN_curl_free_condi}
\bfnabla \times \bfq(\bfr) = 0 .
\] 
by definition of $\bfq(\bfr)$ [Eq.~\eqref{EQN_q_vec_definition}]. This restriction on $\bfq(\bfr)$     plays a central role in determining the low temperature properties of spiral spin liquids.

If spin vortices are taken into consideration, then for any loop (not directly passing through the spin vortex core), one finds
\[
\oint_C \text{d}\bm{l}\cdot \bm{\nabla}\Phi   =  2 \pi n_s,
\]
where $n_s$ is the spin winding number. This implies that a spin vortex at $\bfr'$ is a point source of quantized curl for the momentum vector field,
\[
\bfnabla \times \bfq(\bfr) = 2\pi n_s\delta(\bfr-\bfr').
\]
However, we are mostly concerned with the physics of spiral spin liquids in the absence of spin vortices. In fact, our numerical results   demonstrate that spin vortices are associated with much higher excitation energies than momentum vortices. Hence, from now on, we assume the curl-free condition [Eq.~\eqref{EQN_curl_free_condi}] to be a constraint on the spiral spin liquid system.

\begin{figure}[ht]
\centering
\includegraphics[width=\columnwidth]{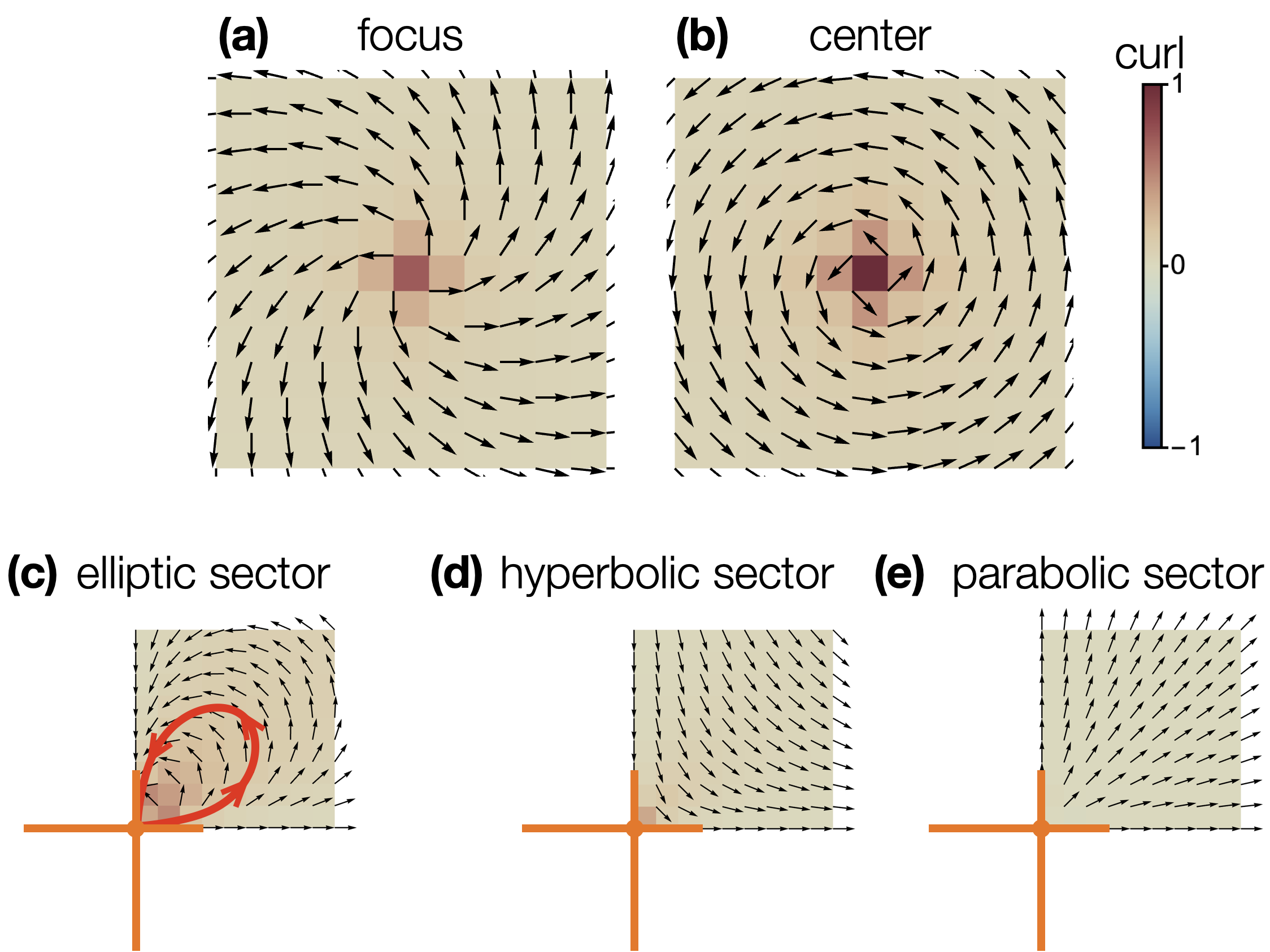}
\caption{Classification of vortices with a background heat map of the curl for the configurations plotted.
(a) A \textit{focus}  around a singularity. It has finite curl and finite divergence at the singularity point. The integrated paths from the vector field converge at the singularity or to infinity.
(b) A \textit{center}  around a singularity. A center is a special limit of focus. It has finite curl at the singularity point and no divergence. The integrated paths from the vector field are exactly  closed loops.
(c) An \textit{elliptic sector} on the top right quarter of the vortex. It has finite curl in the sector.
(d) A  \textit{hyperbolic sector}  on the top right quarter of the vortex. It does not necessarily have  non-zero curl in the sector.
(e) A  \textit{parabolic sector}  on the top right quarter of the vortex. It has  zero curl in the sector.
A vortex is either of class (a), (b),
or has several sectors, each belonging to (c), (d), or (e). 
Vortices of classes focus (a), center (b), or containing  an elliptic sector (c) are forbidden by the curl-free condition.
\label{Fig_vortex_class}}
\end{figure}

The curl-free condition restricts the allowed configurations of the momentum vortices. 
The result and its mathematical deduction  are first summarized below  before elaboration:
\begin{enumerate}
    \item As a starting point, the momentum vector field has to be curl-free [Eq.~\eqref{EQN_curl_free_condi}].
    \item In the neighbourhood of a singular point, the vector field cannot be of the types \textit{focus} [Fig.~\ref{Fig_vortex_class}(a)] or \textit{center} [Fig.~\ref{Fig_vortex_class}(b)].
    It  cannot have an \textit{elliptic}  sector [Fig.~\ref{Fig_vortex_class}(c)] either,
    because all such configurations require finite curl in space.
    The singular point can only have several \textit{hyperbolic} [Fig.~\ref{Fig_vortex_class}(d)] and \textit{parabolic} sectors [Fig.~\ref{Fig_vortex_class}(e)], which can be curl-free. 
    \item  Vortices of winding number $n\ge2$ are forbidden, because they require at least one elliptic sector. 
    Vortices of winding number $n\le 1$ are allowed, but their configuration is still constrained by the condition above.
\end{enumerate}

Two-dimensional vector fields and their vortex singularities
are a well-studied topic in topology  as tangent vector bundle sections in 2D, and also  dynamical systems  \cite{frankel2004geometry,henle1994combinatorial}. 
Around a vortex singularity (also known as critical point in mathematical literature),
the vector field configurations are fully classified.
One possibility is that it can be either a \textit{focus} or a \textit{center} [Fig.~\ref{Fig_vortex_class}(a,b)], surrounding the entire singularity.
The other possibility is that the neighbourhood of the singularity is divided into a few \textit{sectors}. Here, a sector is defined such that at its boundaries the vector field ${\bm q}({\bm r})$ obeys ${\bm q}({\bm r})\parallel {\bm r}$ where the singularity is at ${\bm r}=0$. 
The vector field configuration in each sector is independent from the neighboring ones (as long as continuity on the boundaries is satisfied) and it can be either elliptic, parabolic, or hyperbolic. [Fig.~\ref{Fig_vortex_class}(c,d,e)].

For focus and center configurations the winding number is $n=1$.
For a singularity divided into several sectors, Poincar\'e-Bendixon theorem \cite{frankel2004geometry,henle1994combinatorial} states that the winding number is 
\[
n=1+(e-h)/2,
\label{EQN_Poincare}
\]
where $e$ ($h$) is the number of elliptic (hyperbolic) sectors.
This completes the classification of all vortex configurations with a point singularity.

\begin{figure}[ht]
\centering
\includegraphics[width=\columnwidth]{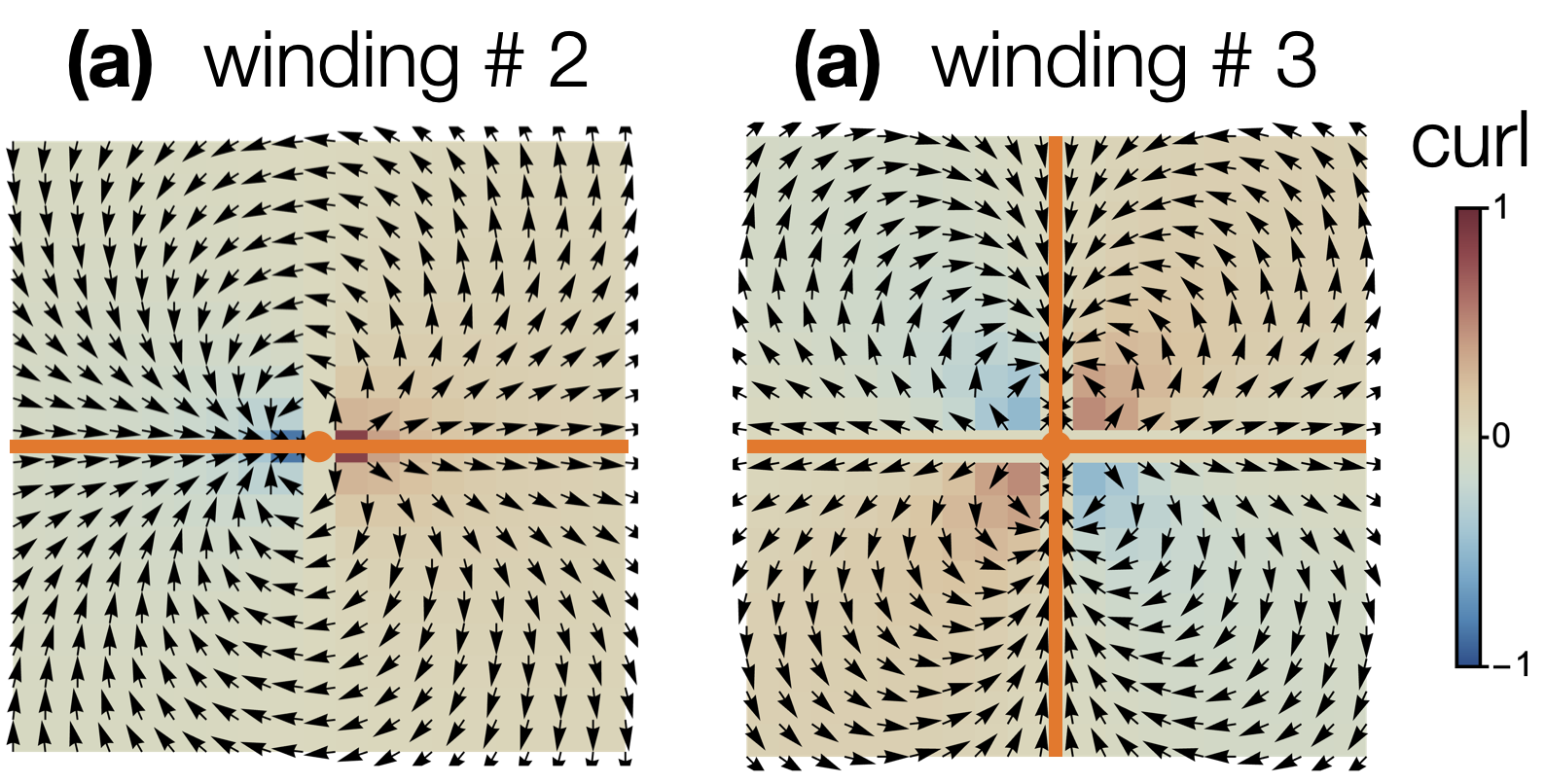}
\caption{Vortices of winding numbers 2 and 3. The winding number 2 vortex has two elliptic sectors divided by the orange lines.
The winding number 3 vortex has four elliptic sectors.
These two examples demonstrate the Poincar\'e-Bendixon theorem  [Eq.~\ref{EQN_Poincare}] and illustrate that they are forbidden in our systems due to their finite curl.
\label{Fig_winding_23}}
\end{figure}

Now we can examine the curl for each class of vortices.
First, it is straightforward to see that focus and center have non-zero curl at the singularity, and hence are excluded in our system.
Second, the elliptic sector has non-zero curl in the entire region.
This can be seen by following the integration path of the vector field [highlighted in red in Fig.~\ref{Fig_vortex_class}(c)]. 
Since such a path forms a closed loop starting and ending at the singularity, and starting and ending vectors  are rotated by a finite angle, we know the enclosed region must have non-zero curl.
In particular, the singular point will have divergent curl.
Hence these sectors are forbidden too. 
Third, the hyperbolic sector may have local non-zero curl but not necessarily, and is therefore allowed in our system. Finally, the parabolic sector is strictly curl-free and allowed.

Now we can conclude on  the vortex restrictions. 
First, since all elliptic sectors are forbidden, and   Poincar\'e-Bendixon theorem states that vortices with winding numbers $n\ge 2$ require at least two elliptic sectors,
such vortices are not allowed (Fig.~\ref{Fig_winding_23}).
Second, the only possible configuration for vortices of winding number $n=1$ is parabolic in the entire neighborhood.
Finally,
vortices of winding number $n\le -1$ are generally allowed, but they cannot have elliptic sectors,
and their hyperbolic sectors still need to be curl-free.

\begin{figure}[htp]
\centering
\includegraphics[width=0.99\columnwidth]{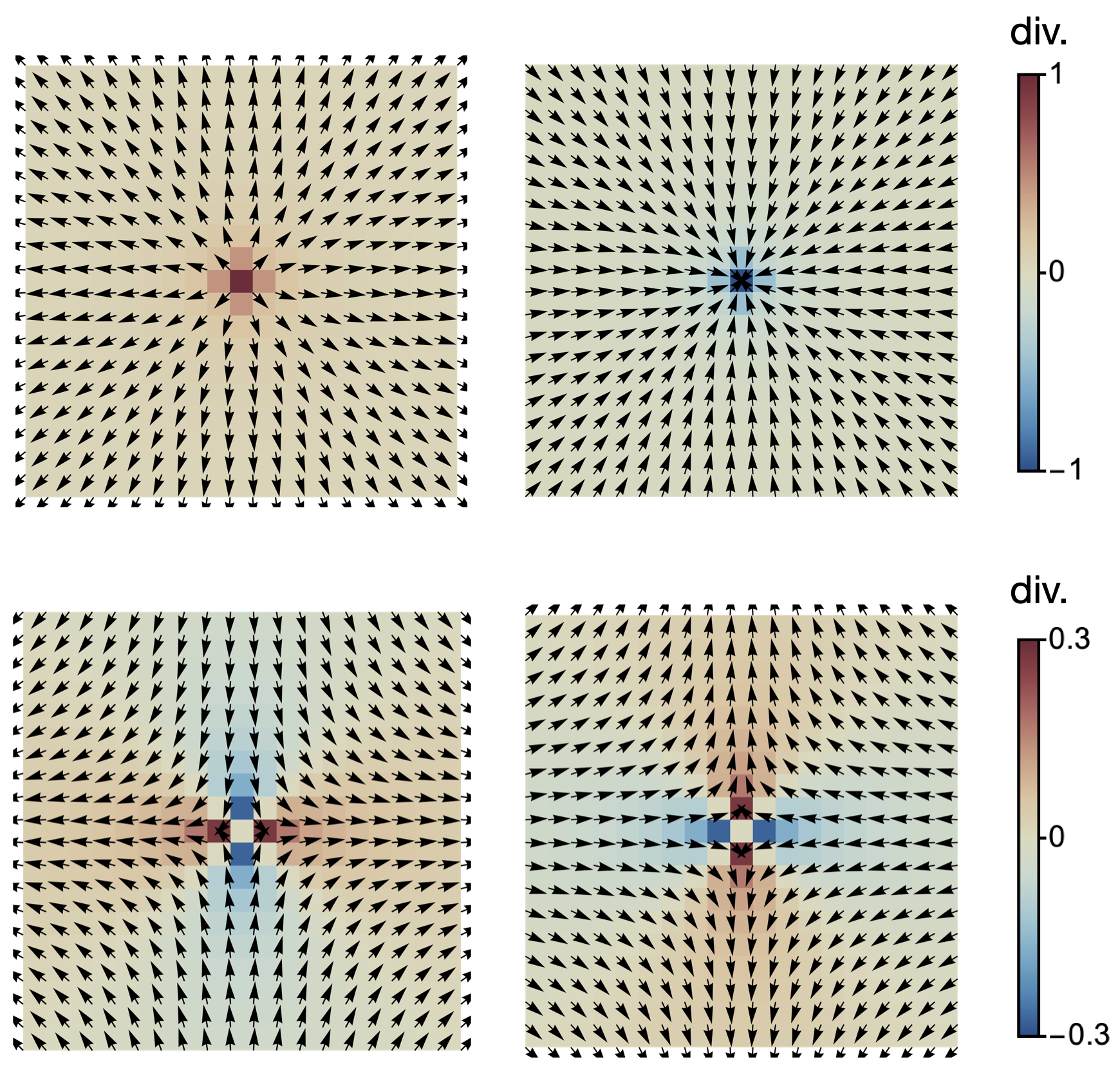}
\caption{Momentum vortices of winding number $1$ (top row) and $-1$ (bottom row), and the heat map of divergence for these configurations.
The winding number $1$ vortices carry a net positive or negative divergence, while the winding number $-1$ vortices carry quadrupoles.}
\label{fig:Fig_div_map_of_vortices}
\end{figure}

Another interesting property is that the absence of curl requires the vortices to carry a finite divergence distribution. 
Here, we briefly discuss the divergence of the simplest vortices with winding numbers $\pm1$, 
since they are most common in our numerical simulations. 
We note that the winding number $1$ vortices carry a net positive or negative, unquantized divergence, as shown in Fig.~\ref{fig:Fig_div_map_of_vortices}.
The two types of winding number $1$ vortices cannot be smoothly transformed into each other.
Interestingly, the winding number $-1$ vortices carry a dominant quadrupole of the divergence distribution (Fig.~\ref{fig:Fig_div_map_of_vortices}). %
They can also carry a net scalar divergence and a dipole, 
but these two quantities can be   smoothly tuned to zero.

\subsection{Hamiltonian in the continuum limit}

\begin{table*}[th]
\caption{Connection between spiral spin liquids and elasticity/rank-2 U(1) gauge theory.
By associating the local momentum vector $\bfq$ in spiral spin liquids with the lattice distortion $\bfu$ in elasticity, various objects in the two types of systems can be identified and further linked to scalar charged rank-2 U(1) gauge theory.
\label{TABLE_spiral_fracton_duality_table}
} 
\begin{tabular}{c @{\hskip 15pt} c @{\hskip 15pt} c} 
	\toprule
	Spiral spin liquid & Elasticity  & rank-2 U(1) theory  \\[3pt] \midrule 
	local momentum ${\bm q}={\bm\nabla}\Phi$	& lattice distortion $\bm{u}$  &  \\[3pt]
	symmetrized Hessian matrix $\mathcal{Q}_{\mu\nu}=\frac{1}{2}(\partial_\mu q_\nu+\partial_\nu q_\mu)$ & strain tensor  $\mathcal{U}_{\mu\nu}=\frac{1}{2}(\partial_\mu u_\nu + \partial_\nu u_\mu$) &   electric field $\epsilon_{\mu\rho}\epsilon_{\nu\sigma}E_{\rho\sigma}$\\ [3pt]
	Hamiltonian [Eq.~\eqref{EQN_continuum}] &  shear modulus Hamiltonian & electric field term \\[3pt]
	spin vortex $\bm{\nabla}\times\bm{q} = 2\pi n_s\delta({\bm r})$, quantized  &  bond angle $\bm{\nabla}\times\bm{u} = \theta$, \textit{periodic}  &  \\[3pt]
	non-existent   & disclination, vortex of bond angle $\theta$,  &  scalar fracton \\[3pt]
	non-existent   & dislocation,   $b_\mu= \epsilon_{\rho\sigma}\partial_\rho \partial_\sigma u_\mu$,  &  dipole of fractons \\[3pt]
	vortex of $\bfq$, winding number $1$   & vacancies/interstitials &  quadruple of fractons, $\text{Tr}\bm{E}$ \\[3pt]
	vortex of $\bfq$, winding number $-1, -2, ...$   & multipoles of vacancies/interstitials  &  higher multipoles of fractons  \\[3pt]
	\bottomrule
	\end{tabular}

\end{table*}

The next ingredient to understand the low-energy behavior of spiral spin liquids is the formulation of a continuum theory.
Assuming that the spins, denoted by their angle $\Phi(\bfr)$, are varying slowly on the lattice, 
we can rewrite the Hamiltonian of Eq.~\eqref{EQN_Ham_SQXY} in the continuum limit.
In the parameter regime of Eq.~\eqref{EQN_parameter_region},  
the Hamiltonian becomes
\begin{equation}
\label{EQN_ham_cont_1}
\begin{split}
&{\cal H}_\text{sq-XY}=\\ 
&\frac{1}{2}\left(- \sum_{\bm{a}=\pm\bm{e}_x,\pm\bm{e}_y}+J_2 \sum_{\bm{a}=\pm\bm{e}_x\pm\bm{e}_y}+\frac{J_2}{2} \sum_{\bm{a}=\pm2\bm{e}_x,\pm2\bm{e}_y}\right)  \\
&\times\sum_{\bm r}\cos(\Phi(\bm{r})-\Phi(\bm{r}+\bm{a})) .
\end{split}
\end{equation}
We then expand $\Phi(\bm{r}+\bm{a})$ in a Taylor series,
\begin{equation}
\label{taylor}
\Phi(\bm{r}+\bm{a})=\sum_{n=0}^\infty\frac{1}{n!}({\bm a}\cdot{\bm\nabla})^n\Phi({\bm r})\;.
\end{equation}  
After inserting Eq.~(\ref{taylor}) into Eq.~(\ref{EQN_ham_cont_1}),
we expand the cosine, keeping   terms which contain up to four derivatives in total, which requires terms up to third orders in Eq.~(\ref{taylor}). 
The continuum limit is performed by replacing $\sum_{\bm{r}}\rightarrow\int d^2\bm{r}$.

Using $\bm{q}=\bm{\nabla}\Phi$,
and denoting  the $2\times2$ Hessian matrix $\bm{\mathcal{Q}}$ as 
\[
\label{EQN_epsilon_def}
\mathcal{Q}_{\mu\nu}  = \partial_\mu \partial_\nu \Phi = \partial_\mu q_\nu,
\]
where Greek indices denote Cartesian coordinates, $\mu,\nu\in\{x,y\}$, we obtain the Hamiltonian written in  $\bfq$ only, with a potential term and a stiffness term:
\begin{equation}
{\cal H}_\text{sq-XY}={\cal H}_p+{\cal H}_s
\end{equation}
where
\begin{align}
{\cal H}_p&=\int \text{d}^2\bm{r} \bigg[ \frac{1+4\delta}{16}q^4 + \delta (-2q^2 + \frac{1}{6}(q_x^4 + q_y^4) )\bigg],\notag\\
{\cal H}_s&=\int \text{d}^2\bm{r} \bigg[ 
 (1+ 4\delta) \mathcal{Q}_{\mu\nu}C_{\mu\nu\rho\sigma}\mathcal{Q}_{\rho\sigma}
+ \frac{\delta}{6} ( \mathcal{Q}_{xx}^2 + \mathcal{Q}_{yy}^2 )\bigg],
\label{EQN_continuum}
\end{align}
with
\[
C_{\mu\nu\rho\sigma} = \frac{1}{16}\left(\delta_{\mu\rho} \delta_{\nu\sigma}+\delta_{\mu \sigma} \delta_{\nu \rho}-\delta_{\mu \nu} \delta_{\rho \sigma}\right) . 
\]  
Here, we have again defined $\delta=J_2-1/4$. Our notation indicates that ${\cal H}_\text{sq-XY}$ consists of two terms, a  potential term ${\cal H}_p$ which determines the energy cost of momentum ${\bm q}$ at each point. The other term ${\cal H}_s$ determines the stiffness  of $\bfq$, i.e. the energy cost of spatially varying spiral configurations.

It is worth noting that in the limit of small $\delta$ {\it and} small momentum $q$, the model becomes rotationally symmetric:
\begin{align}
{\cal H}_p&\rightarrow\int \text{d}^2\bm{r} \bigg( \frac{1}{16}q^4 -2\delta q^2\bigg),\notag\\
{\cal H}_s&\rightarrow\int \text{d}^2\bm{r} \left(
\mathcal{Q}_{\mu\nu}C_{\mu\nu\rho\sigma}\mathcal{Q}_{\rho\sigma}\right),
\end{align}
where we have only kept the leading order in $\delta$ for each power in ${\bm q}$. This form makes it obvious that the potential term acquires a standard symmetric `{\it Mexican hat}' shape.

\subsection{Connection to elasticity and rank-2 U(1) theory}  

Let us now demonstrate the connection between 
our effective field theory of spiral spin liquids and rank-2 U(1) theory/elasticity \cite{XuPRB06,PretkoPRB16,Rasmussen2016arXiv,PretkoPRB17,PretkoPRL18,pretko19,Gromov2019PhysRevLett,Nguyen2020SciPost}.

A spin vortex corresponds to a quantized point singularity of the curl of $\bfq$. 
Let us first assume that they do not appear in the system such that
$\bfq$ is curl-free. Hence, we have  
\[
\partial_\mu q_\nu = \partial_\nu q_\mu
\rightarrow  \mathcal{Q}_{\mu\nu} = \mathcal{Q}_{\mu\nu},
\]
and we can rewrite $\bm{\mathcal{Q}}$  as 
\[
 \mathcal{Q}_{\mu\nu} = \frac{1}{2} \left(\partial_\mu q_\nu + \partial_\nu q_\mu \right) .
 \label{EQN_continuum_Qij}
\] 

We note that by identifying 
the momentum vector $\bfq$ in spiral spin liquids as the lattice distortion in elasticity,
\[
\bfq  \leftrightarrow  \bm{u},
\]
the matrix $\bm{Q}$ in  Eq.~\eqref{EQN_continuum}  becomes the symmetric strain tensor $\bm{\mathcal{U}}$.
\[
\mathcal{Q}_{\mu\nu} \leftrightarrow \mathcal{U}_{\mu\nu} = \frac{1}{2}\left(\partial_\mu u_\nu + \partial_\nu u_\mu \right)
\]
The term $\mathcal{Q}_{\mu\nu}C_{\mu\nu\rho\sigma}\mathcal{Q}_{\rho\sigma}$ in  Eq.~\eqref{EQN_continuum}  is the Hamiltonian of elasticity with  the shear modulus only, and does not  contain  the compression modulus term \cite{Dung2020SciPost}. 

After identifying the strain tensor, 
we can study the correspondence of topological defects in the two systems.
In elasticity,
the bond angle $\bfnabla\times \bfu$ is a \textit{periodic} quantity \cite{pretko19}. 
By definition it corresponds to $\bfnabla\times \bfq$ in spiral spin liquids, which is the spin vortex density and is quantized as an \textit{integer} instead of being periodic.

The fundamental topological defects in elasticity are disclinations, which correspond to a winding of the periodic bond angle \cite{pretko19}. 
Hence, its analog cannot appear in spiral spin liquids, since $\bfnabla\times \bfq$ is not periodic.
Similarly, dislocations $\bm{b}$ as dipoles of disclinations, defined as $b_\mu= \epsilon_{\rho\sigma}\partial_\rho \partial_\sigma u_\mu$, do not appear in spiral spin liquids neither.

It turns out that the winding number $1$ momentum vortices in spiral spin liquids correspond to quadrupoles  of disclinations (or pairs of dislocations), which are vacancies or interstitials (i.e. additional atoms squeezed in the lattice) in elasticity. This is because inserting an additional atom in the lattice makes the other atoms move radially away from it, leading to a lattice distortion field ${\bm u}$ similar to our winding number $1$ momentum vortices, see Fig.~\ref{fig:Fig_div_map_of_vortices}.
These objects have vanishing disclination and dislocation densities.
Instead they are manifested as non-zero divergence of ${\bm u}$ $(\bfq)$, which is also
\[
\bfnabla\cdot \bfq = \Tr{\bm{\mathcal{Q}}}\;.
\]
Since the winding number $1$ momentum vortices cannot have any curl,
the only possibility is to have one parabolic sector in the entire region,
which has divergent divergence at the singularity. Examining the negative winding number vortices in the same way, we find them to carry multipoles of $\bfnabla\cdot \bfq$, i.e., they are multipoles of vacancies or interstitials, see Fig.~\ref{fig:Fig_div_map_of_vortices}.

Utilizing the duality between elasiticy and rank-2 U(1) gauge theory discovered by  Pretko,   Radzihovsky ~\cite{PretkoPRL18}, and   by Gromov ~\cite{Gromov2019PhysRevLett},
we can also establish a connection between spiral spin liquids and rank-2 U(1) theory.
The latter theory is a generalization of Maxwell's electromagnetism, by upgrading the electric field to symmetric tensors, and modifying the definitions of charges, gauge fields, and magnetic fields accordingly. 
A more detailed analysis can be found in Refs.~\cite{XuPRB06,PretkoPRB16,Rasmussen2016arXiv,PretkoPRB17}.
Here we only explain the part relevant to our work.

The mapping between the Hessian matrix in spiral spin liquids and the symmetric tensorial electric field is given by
\[
\mathcal{Q}_{\mu\nu} = \epsilon_{\mu\rho}\epsilon_{\nu\sigma}E_{\rho\sigma},\;
\text{or}\;
E_{\mu\nu} = \epsilon_{\mu\rho}\epsilon_{\nu\sigma} \mathcal{Q}_{\rho\sigma}.\;
\]
To see the existence of a Gauss's law, we notice that for smoothly varying $\bfq$, the scalar charge (fracton) is always zero in the entire space,
\[
\rho\equiv \partial_\mu\partial_\nu E_{\mu\nu} = 
\partial_\mu \partial_\nu \epsilon_{\mu\rho}\epsilon_{\nu\sigma}(\partial_\rho q_\sigma +\partial_\sigma q_\rho)/2 = 0,
\]
where we used $\partial_\mu\partial_\rho \epsilon_{\mu\nu} = 0$.
The fracton scalar charges correspond to disclinations in elasticity.

There is also a conservation law for dipoles of fractons (i.e., dislocations in elasticity):
\[
\begin{split}
&\int \mathrm{d}r^2 x_\mu \rho = \int \mathrm{d}r^2 x_\mu \partial_\rho\partial_\sigma E_{\rho\sigma} \\
=& -\int \mathrm{d}r^2  \partial_\sigma \epsilon_{\mu \alpha}\epsilon_{\sigma \beta}(\partial_\alpha q_\beta+\partial_\beta q_\alpha)/2 = 0\;.
\end{split}
\]
Note that here,  the first term vanishes because 
$\partial_\sigma \epsilon_{\sigma \beta}q_\beta = \bfnabla\times \bfq =0$,
and the second term vanishes because $\epsilon_{\sigma \beta}\partial_\sigma \partial_\beta=0$.
 
Finally, the winding number $1$ momentum vortices are quadrupoles of fractons, and negative winding number momentum vortices are higher multipoles.
A quadrupole of fracton corresponds to vacancies or interstitials in elasticity,
and manifests as non-zero trace of the electric field, or divergence of $\bfq$ (Fig.~\ref{fig:Fig_div_map_of_vortices})
\[
\bfnabla\cdot \bfq = \Tr{\bm{{E}}}\;.
\]

We summarize all the relations between spiral spin liquids, elasticity theory and rank-2 U(1) theory in Table~\ref{TABLE_spiral_fracton_duality_table}.

We have now established the effective theory that sheds more light on the spiral spin liquid nature.
It can be mapped onto classical elasticity (rank-2 U(1) electrostatics) where momentum vortices correspond to multipoles of vacancies or interstitials (fractons). 
These objects are free to move around in the lattice and lead to a spin liquid-like behavior.
This is also consistent with the fact that fractons (disclinations) and dipoles of fractons (dislocations) have restricted mobility, 
while quadrupoles do not.
The correspondence between the Hamiltonians is exact at 
 the critical point 
 \[
J_1 = -1,\; J_2 = 1/4,\; J_3 = 1/8,
\]
but the spiral spin liquid effective theory gains additional terms (which are not of the shear modulus form $\mathcal{Q}_{\mu\nu}C_{\mu\nu\rho\sigma}\mathcal{Q}_{\rho\sigma}$) when $\delta \ne 0$, see Eq.~(\ref{EQN_continuum}). %

\subsection{Numerical results}\label{sec:numerics1}
We now demonstrate how our analytical results on momentum vortices in spiral spin liquids as obtained in the previous subsections manifest in actual numerical simulations of the $J_1$-$J_2$-$J_3$ square lattice XY model. Here, we focus on an intermediate temperature regime, particularly on $T\lesssim T^*$ where $T$ is low enough to enable spiral formation but not too small such that spirals remain liquid due to thermal fluctuations. In the next section, we will investigate the low-temperature rigid vortex network regime.
\begin{figure}[t]
\centering
\includegraphics[width=0.99\linewidth]{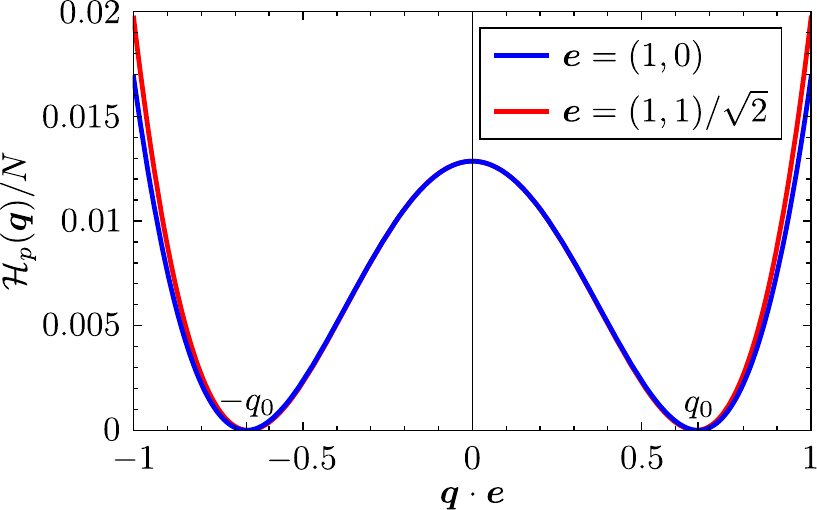}
\caption{Energy per site ${\cal H}_p({\bm q})/N$ [Eq.~(\ref{EQN_continuum})] of homogeneous spiral states with momentum ${\bm q}$, also referred to as potential term in the system's continuum theory. Shown is the Mexican hat shape of this potential for $\delta=0.03$ and two momentum directions, ${\bm q}=(q_x,0)$ or symmetry related directions (blue) and ${\bm q}=(q_x,q_x)/\sqrt{2}$ or symmetry related directions (red).}
\label{fig:mexican_hat}
\end{figure}
\begin{figure}[t]
\centering
\includegraphics[width=0.99\linewidth]{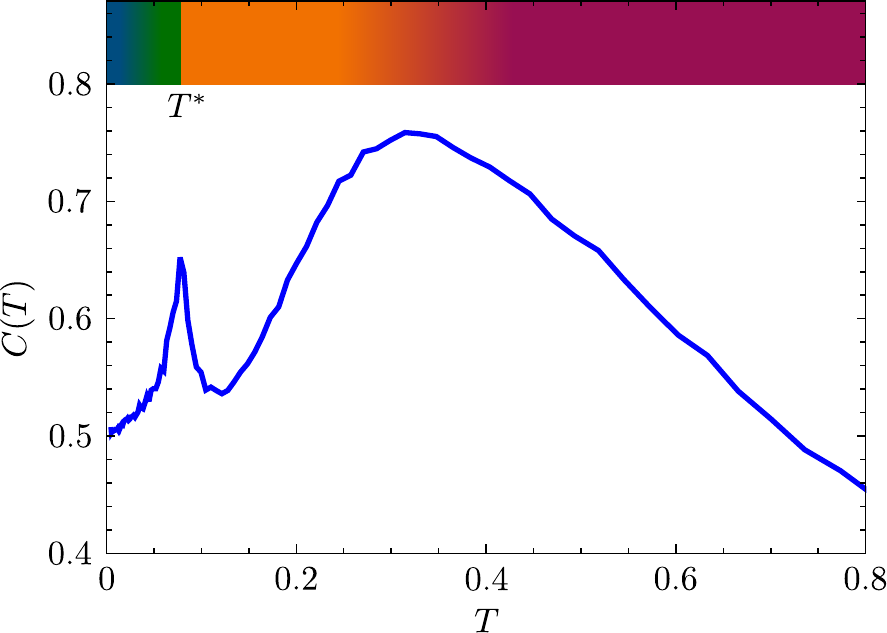}
\caption{Heat capacity per site $C(T)$ as a function of temperature $T$ for $\delta=0.03$. The color bar in the top part of the figure indicates the different temperature regimes with the same color scheme as in the phase diagram of Fig.~\ref{fig1}(b).
\label{fig:heat_cap}}
\end{figure}

Our classical Monte Carlo simulations are based on a standard Metropolis algorithm for a quadratic system with $400\times 400=160000$ sites and periodic boundary conditions. To reduce the autocorrelation times two third of all spin updates are chosen as overrelaxation steps while the other third are regular Monte Carlo updates. One Monte Carlo run includes a total of $6\cdot10^7$ Monte Carlo steps during which the system is cooled down from $T=2$ to $T=0.005$ (in units of $|J_1|=1$) using an exponential cooling protocol. At various selected temperatures, measurements of the energy, heat capacity, spin structure factor, and momentum distribution are performed. Averages are taken over 10 independent Monte Carlo runs.

We choose the parameter $\delta=J_2-1/4=0.03$ in all our simulations below. This value turns out to be suitable for illustrating the physics of momentum vortices, since the potential term ${\cal H}_p$ in Eq.~(\ref{EQN_continuum}) has an almost perfectly rotation symmetric Mexican hat shape (at least in the vicinity of the valley), as shown in  Fig.~\ref{fig:mexican_hat}. For larger $\delta$, the valley of the Mexican hat potential would lose its circular shape, leading to an entropic selection of discrete momenta, which counteracts spiral spin liquid behavior. On the other hand, choosing this parameter too small, the energy gain from spiral formation relative to the ferromagnetic state would become negligible such that spiral states only appear at very low temperatures. For $\delta=0.03$, the classical ground states are spirals with $q_0\approx0.66$ which corresponds to a spiral wave length of $\lambda\approx9.5$ lattice spacings.
\begin{figure}[t]
\centering
\includegraphics[width=0.99\linewidth]{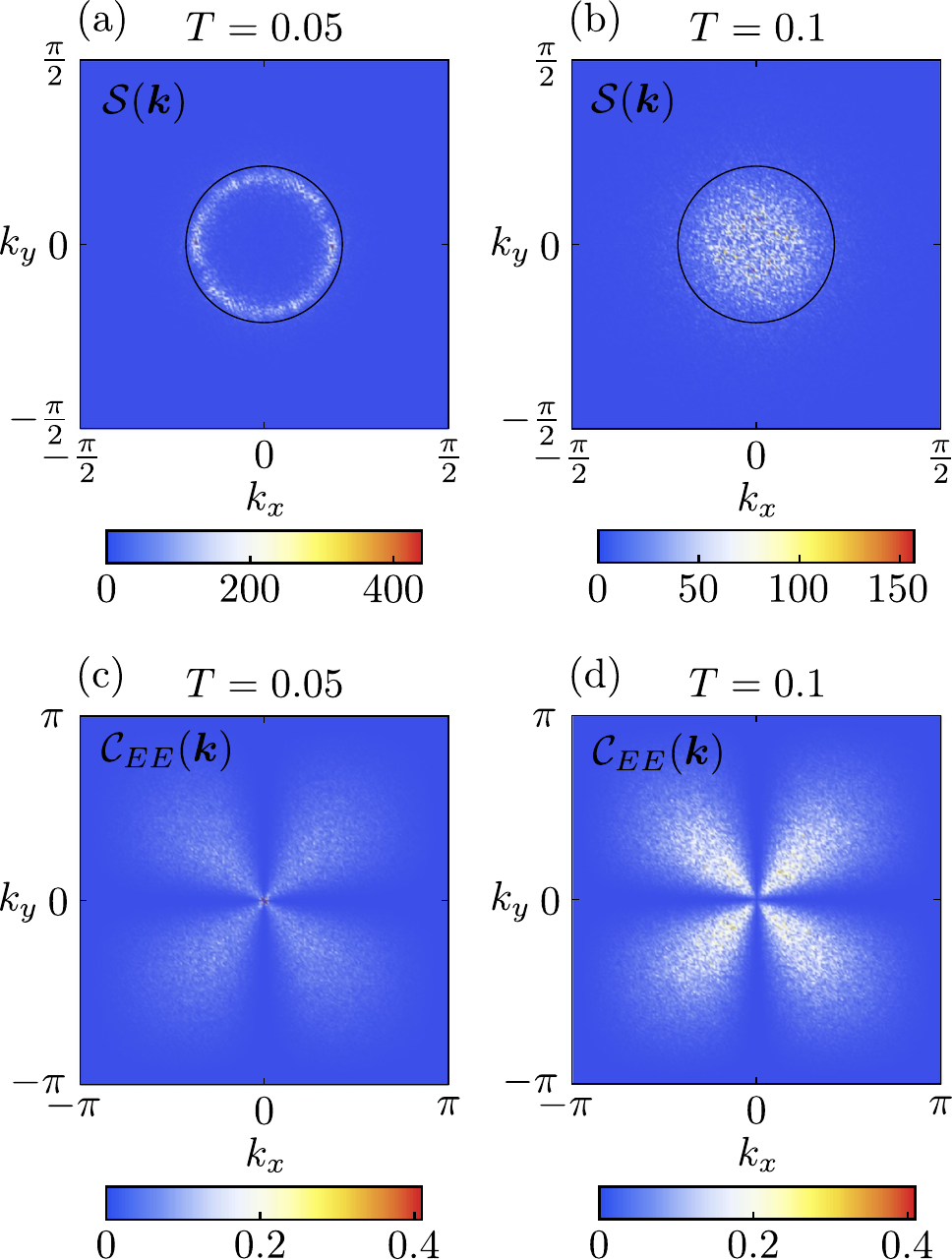}
\caption{Momentum-space properties of spin configurations from classical Monte Carlo at $\delta=0.03$ in the intermediate temperature regime at $T=0.05$ (left column) and $T=0.1$ (right column). (a) and (b) show the spin structure factor ${\cal S}({\bm k})$ [Eq.~(\ref{ssf})] for the states of Figs.~\ref{fig1}(e) and \ref{fig1}(f), respectively. (c) and (d) show the electric field correlator ${\cal C}_{EE}({\bm k})$ [Eq.~(\ref{cee})] obtained by averaging over 10 independent Monte Carlo runs.}
\label{fig:mc_results1}
\end{figure}

Coming from the high-temperature side, the first pronounced feature in the heat capacity (see Fig.~\ref{fig:heat_cap}) is a broad shoulder at $T\approx0.32$ which marks the onset of magnetic correlations. Above this temperature, the system effectively behaves as a paramagnet without any noticeable features in the spin configurations [for an exemplary spin state see Fig.~\ref{fig1}(g) at $T=0.5$].

For temperatures below the broad shoulder the system first enters a regime where correlated spin patterns become visible [see Fig.~\ref{fig1}(f) obtained for $T=0.1$], however, a clear spiral formation does not yet take place. In the spin structure factor ${\cal S}({\bm k})$ defined as
\begin{equation}\label{ssf}
    {\cal S}({\bm k})=\frac{1}{N}\sum_{{\bm r}_i,{\bm r}_j}e^{i{\bm k}({\bm r}_i-{\bm r}_j)}\langle {\bm S}_i\cdot{\bm S}_j\rangle
\end{equation}
this manifests in a broad and featureless peak which roughly fills the area enclosed by the valley of the Mexican hat potential [Fig.~\ref{fig:mc_results1}(b)], hence the name `pancake-liquid' in Ref.~\onlinecite{Shimokawa2019PhysRevLett}. This property indicates that the system can access an extensive manifold of states in momentum space and that in this subspace the potential term ${\cal H}_p({\bm q})$ in Eq.~(\ref{EQN_continuum}) is largely irrelevant. Since the potential term constitutes a key difference to elasticity theory, its irrelevance implies that the system's behavior in this temperature regime is dictated by the analogy to elasticity, i.e., fracton gauge theory. We demonstrate this numerically by plotting the electric field correlation function
\begin{equation}\label{cee}
    {\cal C}_{EE}({\bm k})=\frac{1}{N}\sum_{{\bm r}_i,{\bm r}_j}e^{i{\bm k}({\bm r}_i-{\bm r}_j)}\langle E_{xx}({\bm r}_i)E_{yy}({\bm r}_j)\rangle\;.
\end{equation}
As excepted for a system subject to a generalized Gauss's law $\partial_\mu\partial_\nu E_{\mu\nu}=0$ the electric field correlator needs to obey a specific projector form~\cite{PremPhysRevB.98.165140}, 
\[
\label{EQN_EE_correlator}
\begin{split}
\langle & E_{\mu\nu}(\bm k) E_{\rho\sigma}(-\bm k)\rangle \propto  \\
&\left(\frac{1}{2} \left(
\delta_{\mu \rho} \delta_{\nu \sigma}+\delta_{\mu \sigma} \delta_{\nu \rho}\right)-\frac{k_{\mu} k_{\nu} k_{\rho} k_{\sigma}}{k^{4}}\right)\;,
\end{split}
\]
so that 
\[
{\cal C}_{EE}({\bm k})=\left\langle E_{xx}(\bm k) E_{yy}(-\bm k)\right\rangle \propto  -\frac{k_{x}^2 k_{y}^2  }{k^{4}} 
\]
shows a characteristic fourfold pinch point singularity [Fig.~\ref{fig:mc_results1}(c), (d)].  
The pinch point pattern at $T=0.1$ does not exactly extend to the $\Gamma$ point $\bfq=(0,0)$ which is likely due to a thermal broadening.

Lowering the temperature further, the heat capacity shows a sharp peak at $T=T^*=0.08$ associated with the formation of spin spirals, see Fig.~\ref{fig1}(e). Just below this peak a large density of momentum vortices is observed which are intimately connected with spatial fluctuations of spin spirals, establishing a liquid-like property. Particularly, the spin structure factor now has a ring-like shape indicating that amplitude fluctuations of the spiral momentum are suppressed [Fig.~\ref{fig:mc_results1}(a)]. On the other hand, fluctuations of the direction of spiral momentum are still pronounced as evidenced by the relatively even distribution of signal along the ring. Note that the diameter of the ring is slightly smaller compared to the valley of the Mexican hat potential. This reduction of the momentum amplitude is characteristic for spin configurations with spatially varying direction of ${\bm q}$ which will be further studied in Sec.~\ref{SEC_Eci_with_amp} in the context of domain walls. Since spatially varying momentum directions are the origin of momentum vortices, we interpret the reduction of ring size in Fig.~\ref{fig:mc_results1}(a) as indirect evidence for momentum vortices. The increasing effect of the potential term in the low-energy effective theory slightly reduces the intensity of the fourfold pinch points in Fig.~\ref{fig:mc_results1}(c) but the decreasing influence of thermal fluctuations sharpens the pattern at small ${\bm k}$.

It is worth commenting on the nature of the apparent phase transition at $T=T^*=0.08$. Due to Mermin-Wagner theorem the continuous $U(1)$ spin symmetry cannot be spontaneously broken at a finite temperature; it can only be broken {\it algebraically} which would correspond to a Kosterlitz-Thouless transition. However, as we will argue based on our low-temperature results in Sec.~\ref{sec:numerics2}, we do not observe any indications for such a transition. Spontaneous breaking of discrete time-reversal symmetry is also excluded for a 2D $XY$ model since it acts as $(S_x,S_y)\rightarrow(-S_x,-S_y)$, i.e., it is identical to a $\pi$ rotation in spin space which, however, is a subgroup of the continuous $U(1)$ spin symmetry. In principle, $U(1)$ momentum symmetry could be broken across $T^*$ which would correspond to a spiral selection in the degenerate manifold. Particularly, since this symmetry is not an {\it exact} continuous symmetry on all energy scales but only an approximate one at small energies and small $\delta$, the breaking could even occur at a finite temperature. We studied the behavior of order parameters near $T^*$ which are sensitive with respect to a complete breaking of $U(1)$ momentum symmetry as well as a partial breaking $U(1)\rightarrow\mathds{Z}_2$ and $U(1)\rightarrow\mathds{Z}_4$. However, we do not observe any noticeable features of such order parameters at $T^*$ which also excludes these types of transitions.

We, therefore, conclude that the heat capacity peak at $T^*$ is not related to a real phase transition in the thermodynamic sense but only signals a change of magnetic correlations on finite length scales. Our spin configurations at $T\lesssim T^*$ [Fig.~\ref{fig1}(e)] indicate that in this temperature regime the system forms short range spirals whose momenta ${\bm q}$ are well defined on length scales of a few spiral wave lengths. Beyond this distance spiral momenta show pronounced real-space fluctuations without any long-range patterns, reminiscent of a liquid-like property. The system can gain much energy via this short range spiral formation, as the spin configurations are now  pinned near the valley of the Mexican-hat potential. This sudden energy reduction manifests as a pronounced peak in the heat capacity.

\section{Rigid vortex network phase}
\label{SEC_rigid_fracton_network}

As the temperature decreases, stricter energetic constraints are imposed on the system's spin configurations and the question about the precise nature of energetically optimized momentum vortices arises. 
We first analytically investigate such low-energy momentum vortices via a spiral domain construction where multiple domain walls are radiating from the vortex core. 
Thereafter, we demonstrate that such vortices are indeed found in  low-temperature numerical  simulations.

The overall emerging picture can be summarized as follows: 
At small temperatures,
the Mexican hat potential term $\mathcal{H}_p$ becomes dominant,
and $\bfq$ at every point in space essentially has to lie  exactly on the spiral ring determined by Eq.~\eqref{EQN_GS_q_condition}. 
Furthermore,  the curl-free condition [Eq.~\eqref{EQN_curl_free_condi}] requires all domain walls to be straight lines. 
As a result, 
the system freezes into a rectangular network of winding number $\pm1$ momentum vortices.

\subsection{Vortices without momentum amplitude variation}
\label{SEC_Eci_without_amp}

Let us now consider the case where the local momentum vectors $\bfq(\bfr)$ take values strictly on the spiral momentum ring [Eq.~\eqref{EQN_GS_q_condition}], except of singular lines (domain walls) and points (vortices).  
For simplicity, we  assume that the  spiral momentum ring is an exact circle (as is the case for $\delta\ll1$) 
so it can be parametrized by an angle $\theta$ , 
\[
\label{EQN_q_ring}
\bfq(\bfr) = q_0 (\cos\theta(\bfr) ,\sin\theta(\bfr) ).
\]
Incorporating Eq.~\eqref{EQN_q_ring} in the stiffness part of the Hamiltonian [Eq.~(\ref{EQN_continuum})] and assuming $\delta\ll1$, one obtains 
\[
\mathcal{H}_s \sim  \int \text{d}^2{\bm r}[(\bfnabla q_x)^2 +  (\bfnabla q_y)^2] \sim\int \text{d}^2{\bm r}  [\bfnabla\theta(\bfr)]^2.
\]
Despite a similar structure, this continuum form should not be confused with the  conventional  continuum model of a classical XY magnet (without any spiral degeneracy) featuring a Kosterlitz-Thouless transition.
In our case $\theta(\bfr)$ is still constrained by the   curl-free condition [cf. Eq.~\eqref{EQN_curl_free_condi}],
such that the mechanisms behind the Kosterlitz-Thouless transition do not apply anymore.

Instead, the zero curl condition severely restricts the possible momentum vector field configurations.
As we will show, 
the non-trivial, relevant solutions all have a singularity point (vortex core) and a few straight-line domain walls radiating from it.
These constructions are also motivated by our numerical results which show a rather accurate segmentation of the system into spiral domains, see Fig.~\ref{fig1}(d). 

\subsubsection{Single domain wall}
Let us  first  consider 
a single domain wall that forms a straight line of canting angle $\phi_{12}$ 
between two regions of constant momentum.
In these two regions labeled as $1$ and $2$, we have respectively
\[ \bm{q}_i = q_0(\cos\theta_i,\sin\theta_i),
\quad
i = 1,2.
\]
The curl-free law for the momentum field requires 
the projections of $\bm{q}_{1}$ and $\bm{q}_{2}$ in the direction of the domain wall to be the same, which is 
\[
 \cos(\theta_1 - \phi_{12})
 = 
  \cos(\theta_2 - \phi_{12}).
\]
In terms of the spin textures, this is just requiring the spins on the two sides of the domain wall connect continuously.
The non-trivial solution is
\[
\theta_2   = 2 \phi_{12}-  \theta_1 \;. 
\]
For a given $\theta_1$, the momentum in region $2$ has only one solution.
The corresponding momentum distribution and spin texture is mirror-symmetric regarding the domain wall.
An important corollary from this conclusion is that the domain wall
between two regions of fixed momentum
has to be straight.\\

\subsubsection{Singularity with odd number of discrete domain walls}

\begin{figure}[ht]
\centering
\includegraphics[width=0.5\columnwidth]{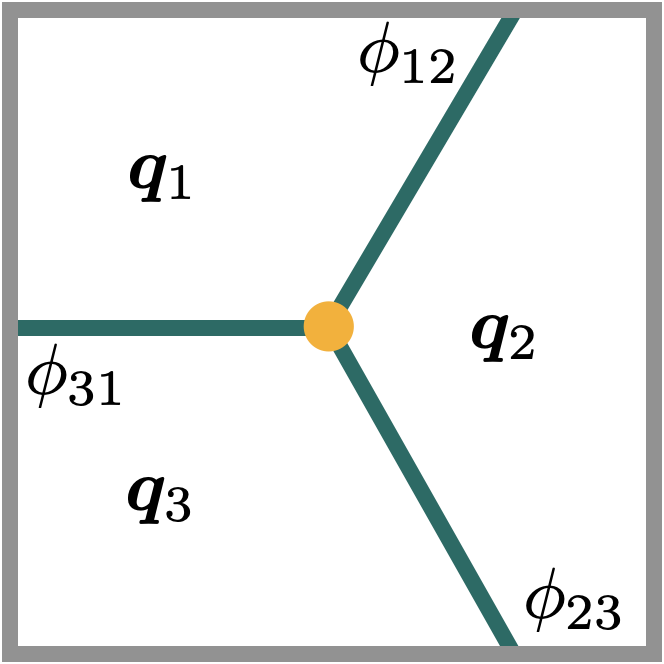}
\caption{Setup of three domains of different momenta and straight line domain walls between them. The $\bfq_i$'s can be solved from Eqs.~(\ref{EQN_3_domain_1}), (\ref{EQN_3_domain_2}), (\ref{EQN_3_domain_3}).
} 
\label{FIG_3_branch_vortex}
\end{figure}

We now consider a singularity -- which will turn out to be a vortex of the momentum vector field -- 
with several domain wall branches 
radiating from it.

We start with the simplest case of a singularity with three branches, as illustrated in Fig.~\ref{FIG_3_branch_vortex}.
The momentum vectors and slopes of domain walls
are as labeled there.

\begin{figure*}[t]
\centering
\includegraphics[width=0.83\textwidth]{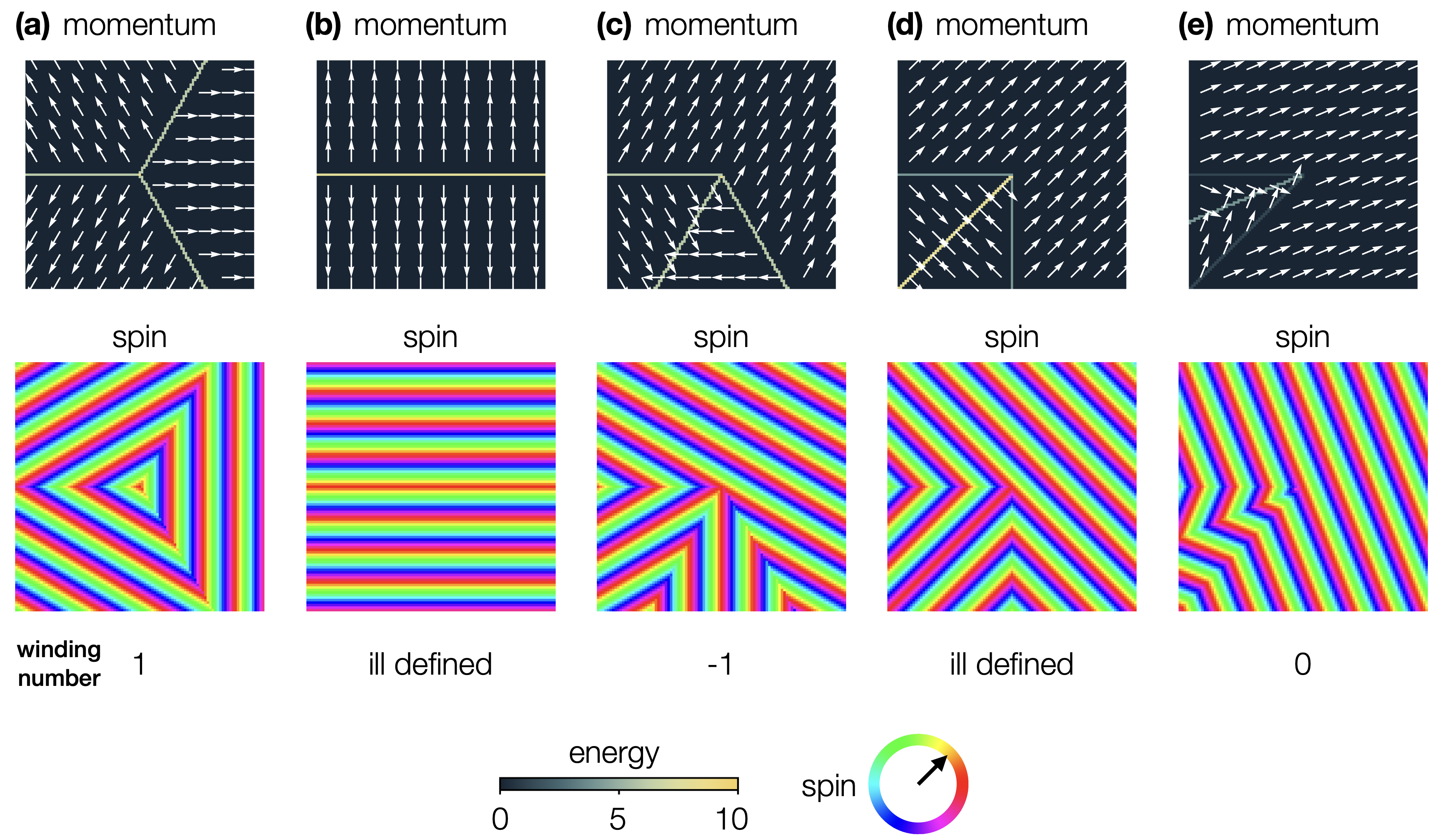}
\caption{Momentum vector field configurations with three domain walls radiating from a singularity.
(a) Momentum vector field and spin texture when the three domain walls are evenly distributed.
Upper panel: Momentum vector field configuration represented by arrows,
on top of a heat map of local energy density.
The momentum vector field forms a vortex of winding number $1$.
Lower panel: spin texture corresponding to the momentum vector field.
(b) Momentum vector field and spin texture when the three domain walls are at angles $0$, $\pi/2$, $\pi$.
The momentum vector field configuration is at a transition point where the winding number is ill-defined.
(c) Momentum vector field and spin texture when the three domain walls are at angles $0$, $\pi/3$, $2\pi/3$.
The momentum vector field configuration forms a vortex of winding number $-1$.
(d) Momentum vector field and spin texture when the three domain walls are at angles $0$, $\pi/4$, $\pi/2$.
The momentum vector field configuration is again at a transition point where the winding number is ill-defined.
(e) Momentum vector field and spin texture when the three domain walls are at angles $0$, $\pi/8$, $\pi/4$.
The momentum vector field configuration forms a vortex of winding number $0$. The energy is in arbitrary units.
} 
\label{Fig_3_domains}
\end{figure*}

The curl-free condition requires 
\begin{align}
\label{EQN_3_domain_1}
\theta_1 + \theta_2 & = 2\phi_{12} + 2n_{12}\pi,\\
\label{EQN_3_domain_2}
\theta_2 + \theta_3 & = 2\phi_{23}+ 2n_{23}\pi, \\
\label{EQN_3_domain_3}
\theta_3 + \theta_1 & = 2\phi_{31} + 2n_{31}\pi ,
\end{align}
whose solutions are
\begin{align}
\theta_1 & = \phi_{12} - \phi_{23}+\phi_{31} +n_{12}\pi -n_{23}\pi +n_{31}\pi  ,\\
\theta_2 & = \phi_{12} + \phi_{23}-\phi_{31} +n_{12}\pi +n_{23}\pi -n_{31}\pi , \\
\theta_3 & = -\phi_{12} + \phi_{23}+\phi_{31} - n_{12}\pi +n_{23}\pi +n_{31}\pi  .
\end{align}
Although $n_{12}, n_{23},n_{31}$ can take any integer values, 
the only two distinct solutions can be obtained by taking $n_{12} = 0,1$ while keeping the other $n_{i,i+1}$ zero.
The two solutions are simply related by globally reversing the directions of all the momentum vectors.

This case can be generalized to all configurations with odd numbers of branches. 
Given $\phi_{i,i+1}$ and $n_{i,i+1}$, finding $\theta_i$ satisfying the curl-free condition is equivalent to solving linear equations
\[
\label{EQN_vector_field_equation}
\begin{pmatrix}
1 & 1 & 0 & \cdots & 0\\
0 & 1 & 1 &\cdots & 0\\
\vdots & \vdots & \vdots & \ddots & \vdots\\
1 & 0 & 0& \cdots &1
\end{pmatrix} 
\begin{pmatrix}
\theta_1  \\
\theta_2 \\
\vdots \\
\theta_m 
\end{pmatrix}
=
\begin{pmatrix}
2\phi_{12}+2n_{12}\pi  \\
2\phi_{23}+2n_{23}\pi \\
\vdots \\
2\phi_{m1}+2n_{m1}\pi
\end{pmatrix}.
\]
The solutions are guaranteed to exist and are unique (up to reversing all momentum vectors), due to the fact that the $m\times m$ matrix
\[
\mathcal{K}_m
\equiv 
\begin{pmatrix}
1 & 1 & 0 & \cdots & 0\\
0 & 1 & 1 &\cdots & 0\\
\vdots & \vdots & \vdots & \ddots & \vdots\\
1 & 0 & 0& \cdots &1
\end{pmatrix} 
\]
always has a non-zero determinant
\[
\det\mathcal{K}_m \ne 0, \quad  \text{for odd $m$.}
\]

We now illustrate the momentum field solutions
for different domain wall distributions.
The cases of three domain walls 
at different angles are shown in Fig.~\ref{Fig_3_domains}.
There, we  plotted the momentum field solutions 
on top of  heat maps of the local energy density,
and also   the spin textures in   separate panels.
In Fig.~\ref{Fig_3_domains}(a),
the three domain walls are evenly distributed,
and the momentum vector field forms a vortex of winding number $1$ with vanishing curl everywhere.
In terms of the spin texture,
the contours of spins form triangle-shaped loops around the center. 

As we squeeze the three domain walls to one side of the plane [Fig.~\ref{Fig_3_domains}(b)], a transition point is reached 
where the winding number becomes ill-defined.
Further narrowing down the angles of domain walls
leads to a state illustrated in Fig.~\ref{Fig_3_domains}(c),
in which the momentum vector field forms a vortex of winding number $-1$.

Another transition point is reached when the three domain walls are squeezed into a quarter of the plane [Fig.~\ref{Fig_3_domains}(d)].
After the transition point, when all three domain walls are confined within an angle of $\pi/2$, the momentum vector field has zero winding number [Fig.~\ref{Fig_3_domains}(e)].

This observation can be generalized to any odd $m$.
When the domain walls are evenly distributed, 
the momentum vector field forms a vortex of winding number $1$.
As one ``squeezes'' the domain walls into a narrower angle, 
the momentum vector field can go through   transitions into vortices of winding numbers $-(m-1)/2$ to $0$.\\

\subsubsection{Singularity with even number of discrete domain walls}
\begin{figure*}[t]
\centering
\includegraphics[width=\textwidth]{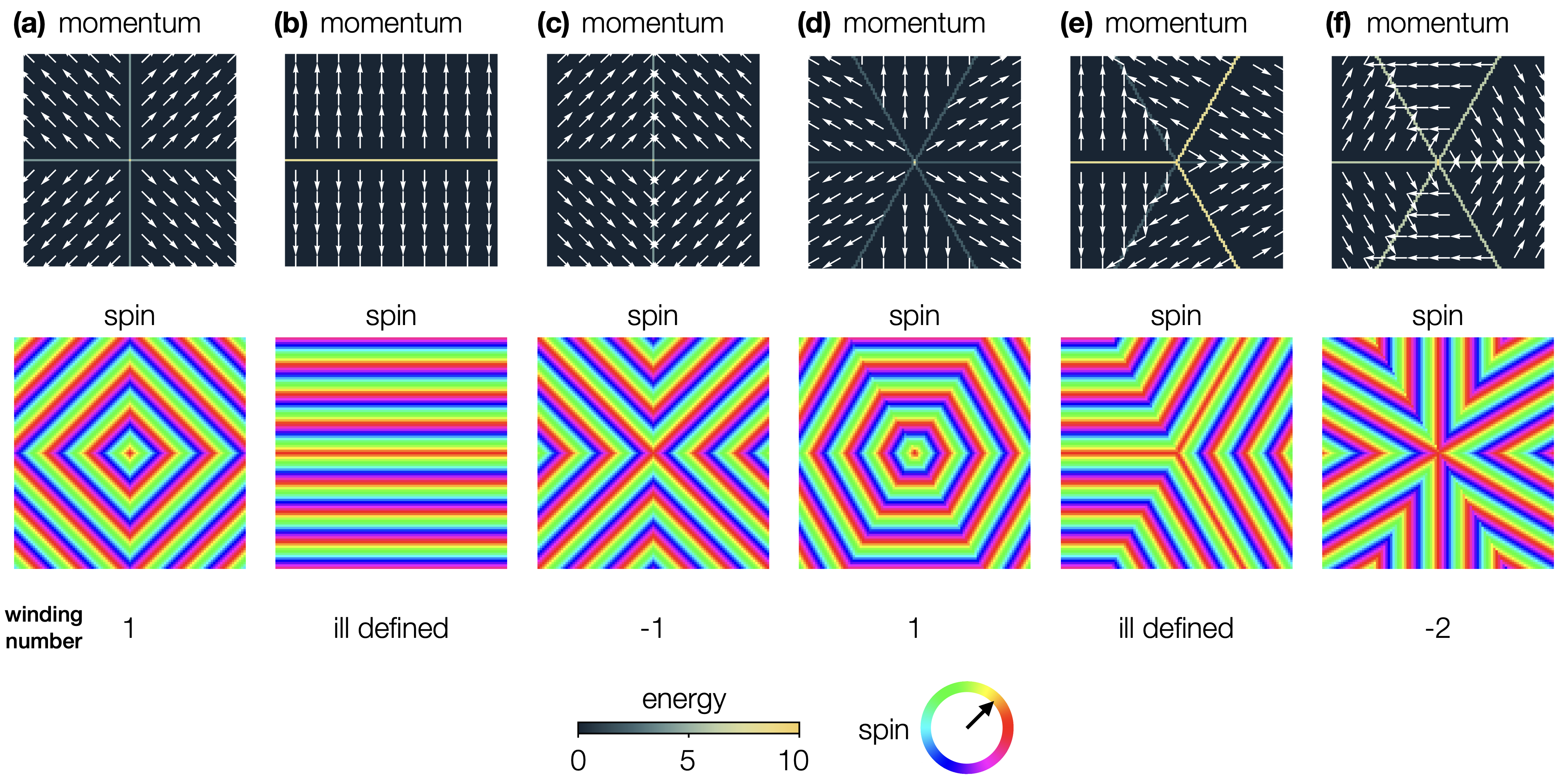}
\caption{Momentum vector field configurations with four and six domain walls radiating from a singularity. Upper panels: Momentum vector field configuration represented by arrows,
on top of a heat map of local energy density. Lower panels: spin texture corresponding to the momentum vector field.
(a-c) Momentum vector field and spin texture when the four domain walls are evenly distributed.
The domain wall angles are fixed but there is a continuous set of solutions for the momentum vector field configuration. 
The momentum vector field forms a vortex of winding number $1$ in (a),  and after going through a transition in (b), the winding number becomes $-1$ in (c).
(b-e) Momentum vector field and spin texture when six domain walls are evenly distributed. 
The momentum vector field forms a vortex of winding number $1$ in (a),  and after going through a transition in (b), the winding number becomes $-2$ in (c).
} 
\label{Fig_4_6_domains}
\end{figure*}

Next we consider a singularity with an even number 
of domain wall branches radiating from it.
Again, determining the momentum vector field is equivalent to solving Eq.~\eqref{EQN_vector_field_equation}.
However, the situation is different from the odd number case, because
\[
\det\mathcal{K}_m = 0, \quad  \text{for even $m$.}
\]
More specifically,  the matrix $\mathcal{K}_m$ has $m-1$ linearly
independent rows.
The last row can be written as linear combination of the first $m-1$ rows
\[
\left(\mathcal{K}_m \right) _{mi}
= \sum_{j=1}^{m-1} (-1)^{j+1}\left(\mathcal{K}_m \right) _{ji}  .
\]
As a consequence, there could be  infinitely many solutions if 
\[
\phi_{12} - \phi_{23} + \cdots  + \phi_{m-1,m} - \phi_{m1} = n \pi,
\label{EQN_even_domain_wall_condition}
\]
and no solution if this condition is not met. 

When there are solutions, 
one can treat $\theta_{m}$
as a free parameter,
and then solve $\theta_{1},\ \dots\ \theta_{m-1}$ from the $m-1$
linearly independent equations.
The solutions then 
form a 1D manifold parametrized by $\theta_m$.
 
For a clearer physical picture, 
we illustrate the cases with four and six domain walls in Fig.~\ref{Fig_4_6_domains} with varying parameter $\theta_m$.
In the case of four domain walls, 
the momentum vector field can form a vortex of winding number $1$ [Fig.~\ref{Fig_4_6_domains}(a)], and after the transition point [Fig.~\ref{Fig_4_6_domains}(b)]
transforms into a  vortex of winding number $-1$ [Fig.~\ref{Fig_4_6_domains}(c)]. 
In the case of six domain walls, 
the momentum vector field can also form a vortex of winding number $1$ [Fig.~\ref{Fig_4_6_domains}(d)],
and  after the transition point [Fig.~\ref{Fig_4_6_domains}(e)]
transforms into a vortex of winding number $-2$ [Fig.~\ref{Fig_4_6_domains}(f)].

The situation can be generalized to any even $m$.
As the parameter $\theta_m$ varies,
the momentum vector field configuration can transit from winding number $1$ to $-(m-2)/2$.

To summarize, 
we have analytically constructed all excitations based on domain walls when $\bfq$ is restricted to be on the spiral ring.
We have found that the momentum vector field 
can form vortices around singular points.
A peculiar feature is that the 
vortices can have any negative winding number $n<0$ and $n=+1$, 
but no winding number $n>1$.
This is a consequence of the curl-free condition [Eq.~\eqref{EQN_curl_free_condi}], which we have already discussed in detail in Sec.~\ref{SEC_winding_number_constraint}.

Since domain walls are associated with an energy cost proportional to their length, all vortices constructed here have the property that their energy cost remains {\it constant} as a function of distance from the core. This is in stark contrast to usual spin vortices where the energy usually decays with $1/r^2$. As is intuitively clear but will be discussed in more detail in Sec.~\ref{SEC_Eci_with_amp}, the excitation energy of a domain wall becomes larger with increasing momentum difference across the domain wall. Consequently, the energetically cheapest vortex with negative winding is the $n=-1$ vortex in Fig.~\ref{Fig_4_6_domains}(c) with four domain walls arranged in angles of $\pi/2$.

In a real physical system, the momentum vector field $\bfq$ is allowed to slightly deviate away from the ring associated with a `potential energy cost'. 
This lead to a softening of the sharp domain walls, 
and will be studied in detail in Sec.~\ref{SEC_Eci_with_amp}.
However, the key physics discovered here -- that the low-energy excitations are vortices of winding number smaller than or equal to $1$ -- is preserved, 
and plays a major role in determining the low energy physics of the spiral spin liquid.

\subsubsection{Smooth vortices}

For the sake of completeness, here we treat the case where the vector field ${\bm q}$ is analytic everywhere except of the vortex core at ${\bm r}=0$. This will result in a smooth vortex with winding number $n=1$ [see Fig.~\ref{fig:vortices_examples} (left)], which does not have the sharp domain walls discussed above. This case can also be understood as the limit of infinitely many domain walls, densely covering the entire plane.
However, because all negative winding number vortices are made of a singularity and several straight line domain walls,
the smooth winding number $n=1$ vortex is actually not observed in numerical simulations due to its incompatibility with $n\leq1$ vortices.

Let us  denote the momentum field   as  $\bm{q}(\bm{r})=\bm{q}(\theta(\bm{r}))$ where the polar angle $\theta=\theta(\bm{r})$ defines the inplane orientation of $\bm{q}$, see Eq.~(\ref{EQN_q_ring}).
The curl-free condition [Eq.~\eqref{EQN_curl_free_condi}]  then becomes
\begin{equation}
(\bm{\nabla}\theta)\times\partial_\theta \bm{q}=0\;.\label{EQN_curlfree2}
\end{equation} 
Since we are constructing excited states, we first exclude the trivial case $\bm{\nabla}\theta=0$. 
Notice that $\partial_\theta \bm{q}\perp\bm{q}$, therefore
the condition in Eq.~(\ref{EQN_curlfree2}) implies that 
\[
\bm{\nabla}\theta\perp \bm{q}
\]
must hold. 
Furthermore, at each point $\bm{r}_0\neq0$ the gradient $\bm{\nabla}\theta(\bm{r}_0)$ is oriented perpendicular to the contour line of constant $\theta(\bm{r})$ running through $\bm{r}_0$. 
This means that the momentum field $\bm{q}(\bm{r})$ aligns in the direction of the contour line of constant $\theta(\bm{r})$. 
It follows immediately that contours of constant $\theta(\bm{r})$ must be straight lines; if they are bent, $\bm{q}$ would have different orientations on different points of the contour, against the property that the contour has constant $\theta(\bm{r})$.

In a field $\theta({\bm r})$ with the property that lines of constant $\theta$ are straight, a singularity at ${\bm r}=0$ implies that these lines cross at ${\bm r}=0$. In other words, lines of constant $\theta(\bm{r})$ point radially away from $\bm{r}=0$ which means $\bm{q}(\bm{r})\sim \bm{e}_r$, where $\bm{e}_r$ is the radial unit vector.
Including the normalization of the momentum yields the two solutions
\begin{equation}
\bm{q}(\bm{r})=\pm q_0\bm{e}_r\;.\label{single_vortex}
\end{equation}
One solution is illustrated in Fig.~\ref{fig:vortices_examples}.
These states correspond to vortices in momentum space with a phase winding of $n=+1$, as one would intuitively expect.
Our proof shows that they are the only viable states.

By taking this configuration as a given input state and calculating its classical energy in a spiral spin liquid phase, 
we find that its energy decays as $1/r^2$, 
just like  for a usual spin vortex. 
Note, however, that for a spin vortex, there is the freedom to perform a global $U(1)$ rotation of all spins. 
This is different for a momentum vortex with $n=+1$ which only allows for the $\mathds{Z}_2$ transformation of globally inverting the momentum, $\bm{q}\rightarrow-\bm{q}$, leading to the two signs in Eq.~(\ref{single_vortex}).

\begin{figure}[ht]
\centering
\includegraphics[width=0.8\columnwidth]{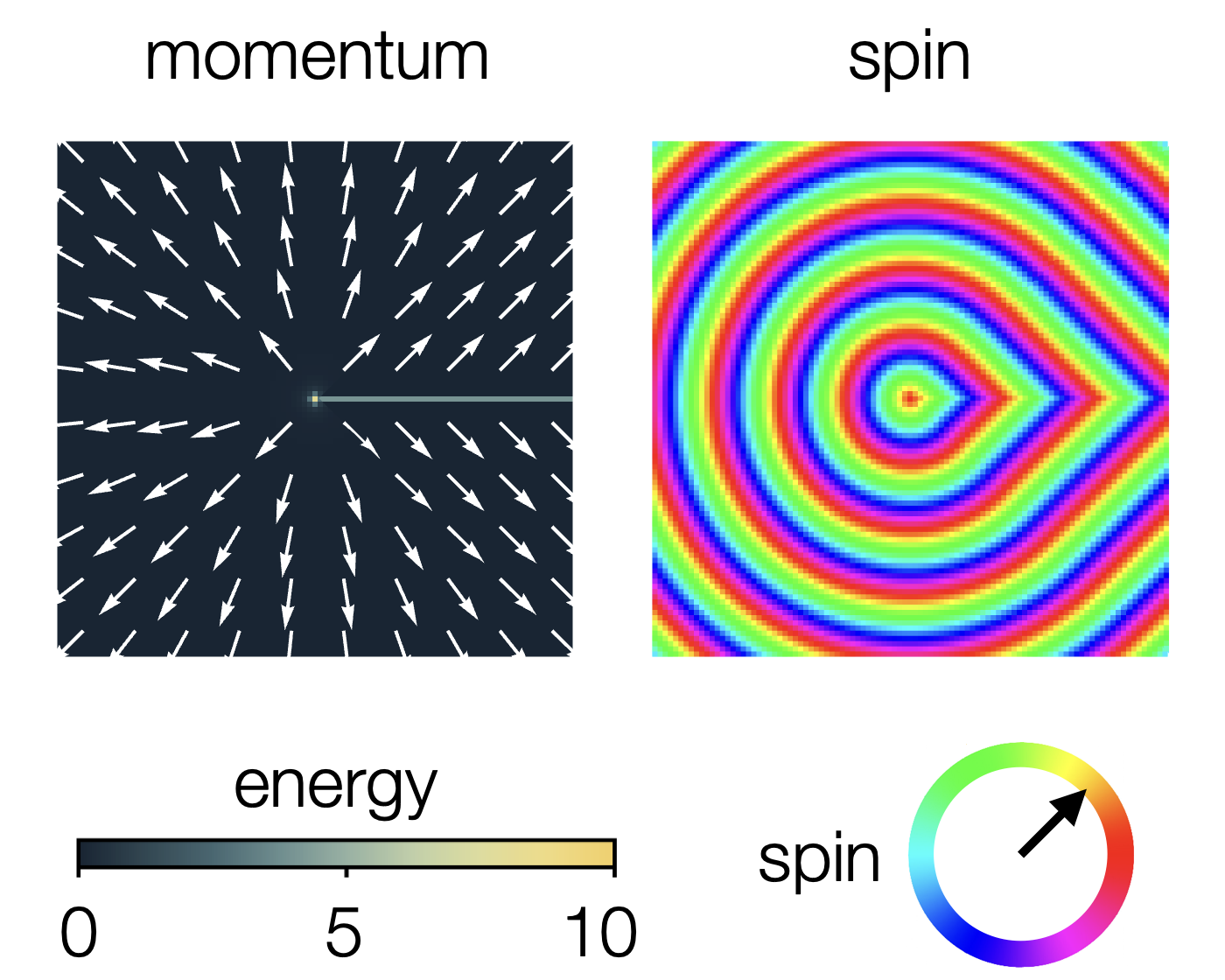}
\caption{
A hybrid momentum vector field configuration constructed by gluing parts of  Fig.~\ref{fig:vortices_examples}(left) and Fig.~\ref{Fig_4_6_domains}(a) together.}
\label{Fig_mix_domain_wall}
\end{figure}

Finally, we mention briefly that proper  mixtures of smooth and discrete domain wall configurations can also be constructed.
For any discrete domain wall solution,
if two radial lines of momentum match with the solution of smooth domain walls
[Eq.~\eqref{single_vortex}], 
the two solutions can be cut and pasted together to form a new solution. 
An example is given in Fig.~\ref{Fig_mix_domain_wall},
where we take the right quarter of Fig.~\ref{Fig_4_6_domains}(a) and the rest of Fig.~\ref{fig:vortices_examples} (left). However, for the same reason of compatibility with negative winding number vortices,
these hybrid vortices do not actually occur in numerical situations.

\subsection{Formation of rigid vortex network}

Let us now zoom out to the entire lattice, and argue the formation of a rigid vortex network. 

First, vortices  and straight line domain walls must form a network.
Assuming periodic boundary conditions,
the entire lattice will have vortices of both positive winding number, which can only be $1$, and negative winding numbers.
The negative winding number vortices do not have a smooth configuration -- they must have several branches of straight,
sharp domain walls radiating from the vortex cores.
For this reason,
the positive winding number vortices cannot have smooth configurations neither,
as they are incompatible with domain walls from the negative winding number vortices.
Hence, 
all vortices should have only straight-line domain walls which interconnect into a network.
In this network,
the edges (links) are the domain walls,
and the nodes (vertices) are the vortices.

Furthermore, 
the configurations of vortices and domain walls that can actually appear in the system are selected by their energy cost, especially at very low temperatures.
Numerically,
we find that vortices with four   perpendicular domain walls are the energetically most favorable ones. 
This property results from a balance between positive and negative winding number vortices: 
while winding number $1$ vortices usually prefer more domain walls to lower their energy,
this does not always apply to negative winding number vortices.
Therefore, a rectangular network of vortices and domain walls is expected at low temperatures.

The vortex network is  rigid. 
This can be deduced from the analytical solutions for a given configuration of domain walls [Eq.~\eqref{EQN_vector_field_equation}].
Note that, 
around a vortex, 
the four domain wall angles must satisfy Eq.~\eqref{EQN_even_domain_wall_condition} in order for a momentum vector field solution to exist.
If we move the positions of one or few vortices, then for some of the neighboring vortices, 
only one of the domain wall angles will change, which obviously will violate the condition in Eq.~\eqref{EQN_even_domain_wall_condition}.
The movement of domain walls or vortices must be of sub-system type at least.
Hence,  the vortex network  becomes highly rigid at low temperatures, when the local momentum vectors are almost strictly on the spiral ring and Eq.~\eqref{EQN_even_domain_wall_condition} is enforced.  

In a system with a frozen rigid vortex network,
it becomes very difficult for the vortices to annihilate each other in order to further reduce the system's energy, since any movement of vortices requires a global, cooperative change of the spin configuration. 

These arguments imply that Kosterlitz-Thouless-type behavior is not expected in our system. 
Fundamentally, this is because of the curl-free condition [Eq.~\eqref{EQN_curl_free_condi}] and the appearance of vortices that break the momentum vector U(1) rotational symmetry.
Note that, although the ground state manifold has such a degeneracy/symmetry, 
the excitations (momentum vortices) do not. 
More specifically,
this leads to the formation of straight line domain walls, 
with an energy cost proportional to their {\it length}, in  stark contrast to usual spin  vortices.
These differences significantly control  the thermodynamic behavior of the rigid vortex  network phase,
and make the Kosterlitz-Thouless phase transition physics inapplicable.

\subsection{Domain wall broadening}
\label{SEC_Eci_with_amp}

To make sure that our analysis in the previous section is valid, we still have to address the question whether our assumption that the momentum vector is strictly confined to the spiral ring actually applies.
We will see in this section that,
although sharp domain walls do not represent the exact energetically optimal spin configurations in a real system,
reintroducing amplitude variations  only broadens the domain walls to a finite extend,  at a scale much smaller  compared to the rigid vortex network.
Consequently, the sharp domain wall assumption can effectively be considered as valid.

Let us now study the finite broadening of domain walls that happens in  more realistic settings, when the momentum vector field $\bfq$ is allowed to fluctuate away from the spiral ring.
Particularly, we will demonstrate that an energy-optimized broadened domain wall has a simple analytic description in the continuum limit. 
We will calculate the ideal form of the rounded domain wall and show that it comes with a low energy cost if the momentum difference across the domain wall is small. 
Note that this rounding is necessarily associated with amplitude variations of the momentum.  
\begin{figure}[t]
\centering
\includegraphics[width=0.95\linewidth]{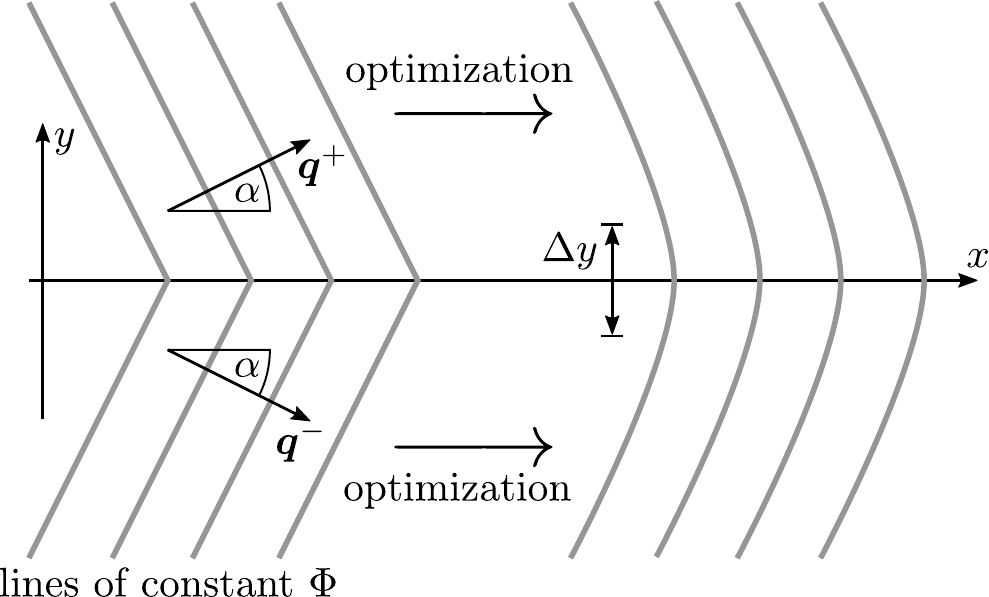}
\caption{Optimization of a domain wall: A discontinuous, non-optimized domain wall (left) characterized by an angle $\alpha$ consists of two homogeneous ground state spin spirals with momenta ${\bm q}^+$ and ${\bm q}^-$ arranged symmetrically around the discontinuous line. Upon optimization, lines of constant spin directions $\Phi$ become rounded and the domain wall acquires a finite width $\Delta y$ (right).}
\label{fig:optimize}
\end{figure}

We first specify the type of spin configuration that we will describe with the help of our continuum model. Let us start with a discontinuous (non-optimized) horizontal domain wall along the $x$ axis which separates two homogeneous spin spirals with momenta ${\bm q}^+$ and ${\bm q}^-$ in the upper and lower half planes, respectively. Furthermore, we require that both spirals are ground states, i.e., $|{\bm q}^+|=|{\bm q}^-|=q_0$, where the ground state momentum $q_0$ follows from minimizing the potential term $\mathcal{H}_p$ in Eq.~(\ref{EQN_continuum}) in the limit $\delta\ll1$, yielding $q_0=4\sqrt{\delta}$. Since the spin configuration must be symmetric around the $x$ axis due to the curl-free condition (see Fig.~\ref{fig:optimize}),
the Cartesian components obey $q^+_x=q^-_x$ and $q^+_y=-q^-_y$.  
In the following, we will characterize the domain wall by $\alpha\in[0,\pi/2]$ which is the angle enclosed by the $x$ axis and ${\bm q}^+$ (or ${\bm q}^-$) such that
\begin{align}
q^\pm_x&=q_0\cos\alpha=4\sqrt{\delta}\cos\alpha\;,\notag\\
q^\pm_y&=\pm q_0\sin\alpha=\pm 4\sqrt{\delta}\sin\alpha\;.\label{qpm}
\end{align}
Without loss of generality we have fixed the sign $q^+_y>0$ in this equation. With these definitions, $\alpha$ describes the strength of the discontinuity where $\alpha=0$ stands for no domain wall and $\alpha=\pi/2$ corresponds to the maximal momentum jump across the domain wall.

The above spin configuration is an excitation where all the energy cost is concentrated  along the infinitely narrow domain wall. 
Upon optimization the energy will spread out into a strip of finite width, and contours of constant spin angle $\Phi$ become rounded (see Fig.~\ref{fig:optimize}). 
In the following, we will use the continuum model in Eq.~(\ref{EQN_continuum}) to calculate the ideal momentum distribution ${\bm q}(\bm r)=(q_x({\bm r}),q_y({\bm r}))$ of this state. Most importantly, the optimized momentum configuration is still translation invariant along the $x$ axes, such that lines of constant $\Phi$ transform into each other by parallel shifts along the $x$ axes. As a result, the functional dependencies of ${\bm q}({\bm r})$ reduce to
\begin{equation}
{\bm q}({\bm r})\equiv{\bm q}(y)=(q_x,q_y(y))\;,
\end{equation}
which means that $q_x$ is constant across the entire spin configuration and $q_y$ is only a function of $y$ such that the optimization becomes an effective 1D problem. 
This also guarantees that the curl-free condition [Eq.~\eqref{EQN_curl_free_condi}] is not violated in the optimization process.
Using this property to simplify the continuum model in Eq.~(\ref{EQN_continuum}) and exploiting Eq.~(\ref{qpm}) one finds that the energy $E$ per length $l$ of the domain wall in leading order in $\delta$ is given by
\begin{align}
\frac{E}{l}&=\frac{1}{16}\int_{-\infty}^{+\infty}dy\left[-2\left(q_y^+\right)^2 q_y^2+q_y^4+(\partial_y q_y)^2\right]\notag\\
&=\int_{-\infty}^{+\infty}dy\mathcal{L}\;.\label{domain_functional}
\end{align}
Here, terms constant in $y$ are neglected (these terms, however, may still have an $\alpha$ dependence). This functional needs to be minimized with respect to $q_y(y)$ where the boundary conditions follow from the fact that far away from the domain wall the spin configurations are given by the initial homogeneous spiral states with momenta ${\bm q}^+$ and ${\bm q}^-$:
\begin{equation}
\lim_{y\rightarrow\pm \infty}q_y(y)=\pm q_y^+\;.\label{boundaries}
\end{equation}
Using Euler-Lagrange equation
\begin{equation}
\partial_{q_y}\mathcal{L}=\partial_y\partial_{\partial_y q_y}\mathcal{L}
\end{equation}
this leads to the differential equation
\begin{equation}
\partial_y^2 q_y=-2\left(q_y^+\right)^2 q_y+2q_y^3\;.
\end{equation}
The solution respecting the boundary conditions in Eq.~(\ref{boundaries}) is found to have a simple form:
\begin{equation}
q_y(q)=q_y^+\tanh(q_y^+ y)\;.
\end{equation}

It is worth highlighting two properties of this result. First, from the argument of the hyperbolic tangent, the width $\Delta y$ of the optimized domain wall is given by
\begin{equation}
\Delta y=\frac{1}{q_y^+}=\frac{1}{4\sqrt{\delta}\sin \alpha}\;,\label{deltay}
\end{equation}
which indicates that domain walls with small $\alpha$ have a diverging width. On the other hand, amplitude variations of the momentum are small in this limit such that, in total, these domain walls are still energetically cheap. For a typical angle $\alpha=\pi/4$ which is realized for the domain walls of momentum antivortices with the lowest possible energy one obtains $\Delta y=1/\sqrt{8\delta}$. Inserting $\delta=0.03$ as used in our numerics yields a rather small width of $\Delta y\approx2$ lattice constants. [It is, of course, questionable whether our continuum model where Eq.~(\ref{taylor}) is truncated still approximates the system reasonably well when momentum variations occur on such short distances. Better estimates are expected for smaller $\delta$.] Our numerics results below confirm that domain walls are narrow with a widths of only a few lattice spacings. 

Second, it is instructive to calculate the maximal excitation energies in the center of a domain wall at $y=0$. To this end, we compare the Lagrangian $\mathcal{L}$ in Eq.~(\ref{domain_functional}) (which provides the energy per unit area, or equivalently, per site) at $y=0$ (where $q_y=0$ and $\partial_y q_y=\left(q_y^+\right)^2$) and at $y\rightarrow\infty$ (where $q_y=q_y^+$ and $\partial_y q_y=0$). The maximal excitation energy per site of a domain wall with angle $\alpha$ then reads as 
\begin{equation}
\Delta E_\alpha=\frac{\left(q_y^+\right)^4}{8}=32\delta^2\sin^4\alpha\;.
\end{equation}
The scaling $\Delta E_\alpha\sim\alpha^4$, which holds for $\alpha\ll 1$, implies remarkably small excitation energies in the limit of vanishing $\alpha$. One may conclude that the formation of domain walls with small $\alpha$ provides a natural source for low-energy thermal fluctuations.

For a typical domain wall with $\alpha=\pi/4$ the excitation energy is given by $\Delta E_{\pi/4}=8\delta^2$. Comparing this value with the local excitation energy of a trivial ferromagnetic state, $\Delta E_{\text{fm}}=16\delta^2$, which defines another characteristic energy scale of the system, one realizes that $\Delta E_{\pi/4}$ cannot be considered small. The fact that such domain walls are still observed at small temperatures (see below) is related to the fact that this energy is confined to a narrow strip.

One may conclude that this confinement of energy represents the key property of domain walls and implies that they remain well defined and narrow even far away from vortices.

\subsection{Numerical results}\label{sec:numerics2}
\begin{figure}[t]
\centering
\includegraphics[width=0.99\linewidth]{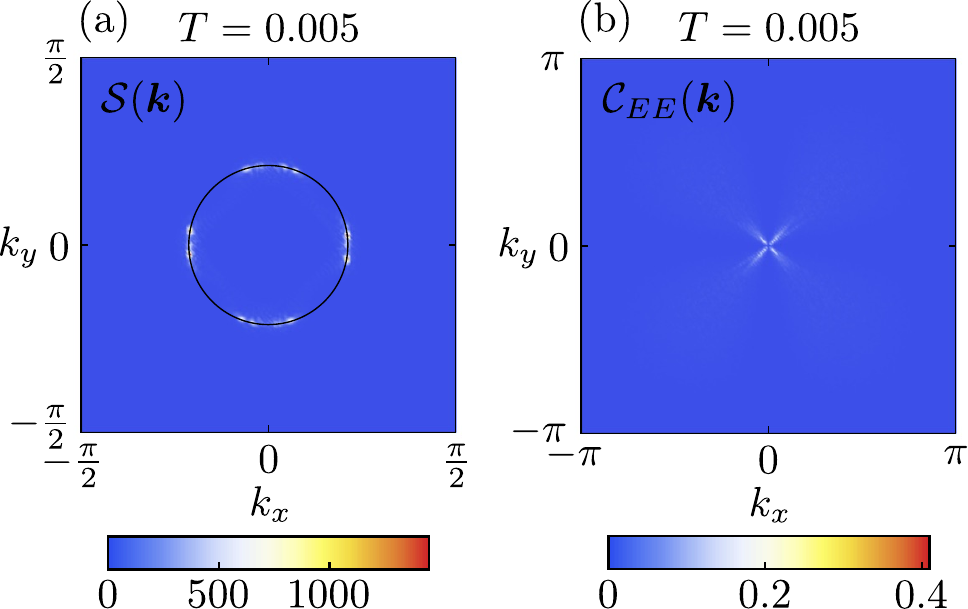}
\caption{Momentum-space properties of spin configurations from classical Monte Carlo with $\delta=0.03$ in the low-temperature regime at $T=0.005$. (a) shows the spin structure factor ${\cal S}({\bm k})$ [Eq.~(\ref{ssf})] for the state in Fig.~\ref{fig1}(d). (b) shows the electric field correlator ${\cal C}_{EE}({\bm k})$ [Eq.~(\ref{cee})] obtained by averaging over 10 independent Monte Carlo runs.}
\label{fig:mc_results2}
\end{figure}

Having discussed the physical properties of energy-optimized momentum vortices and their domain wall construction we now study their occurrence in our low-temperature Monte-Carlo simulations at $\delta=0.03$.

Further decreasing the temperature in the spiral spin liquid regime investigated in Sec.~\ref{sec:numerics1}, the heat capacity (Fig.~\ref{fig:heat_cap}) remains completely featureless below $T^*=0.08$. This indicates that all changes which a spiral spin liquid undergoes when cooling it down are smooth crossovers while sharp phase transitions are not observed. Note that in the low temperature limit, the heat capacity approaches $C=1/2$, as is expected for a XY model with one quadratic mode per site.

The most obvious qualitative change between $T=0.05$ and $T=0.005$ in Figs.~\ref{fig1}(d) and \ref{fig1}(e) is that the spiral domains become well-defined and the domain walls straighten, turning into a network of rigid lines running through the entire system. The intersections of domain walls define the locations of momentum vortices which have winding numbers $n=1$ or $n=-1$. Note that in agreement with our analytical investigation of energy optimized spin configurations, all vortices with $n=-1$ have four domain walls radiating from the center with angles close to $\pi/2$ between them. As discussed above, $n=1$ vortices could in principle be constructed in a smooth way without domain walls, see Fig.~\ref{fig:vortices_examples}. However, due to their incompatibility with $n=-1$ vortices we observe that in most of our numerical outputs $n=1$ vortices adapt the geometry of $n=-1$ vortices and, likewise, show four domain walls radiating from the core.

No binding effects between vortices and antivortices are observed; on the contrary, with decreasing temperature, their distance increases. We, hence, exclude the possibility of a Kosterlitz-Thouless transition resulting from U(1) momentum symmetry. The same applies to spin vortices which are occasionally seen in the numerical outputs, even at the lowest simulated temperatures, see encircled spin configurations in Fig.~\ref{fig1}(d) (in the spin pattern of an ideal spiral state, a spin vortex has a similar shape as a dislocation in a regular crystal). None of our independent Monte Carlo runs reveals a binding into spin vortex-antivortex pairs.

In most of our low-temperature spin configurations from Monte Carlo, the energetic preference of $n=-1$ vortices with four domain walls leads to regular square-shaped patterns that spread over the entire lattice. This goes long with the selection of four momenta along the spiral ring, as is evident in the spin structure factor at $T=0.005$, see Fig.~\ref{fig:mc_results2}(a). In the case of a perfect circular symmetric potential term $\mathcal{H}_p$, these four momenta could also occur at any rotated positions on the spiral ring. However, the small deviations from rotation symmetry at finite $\delta$ pin them along the cartesian $q_x$ and $q_y$ axes in Fig.~\ref{fig:mc_results2}(a). For this reason, a finite-temperature spontaneous breaking of $U(1)$ momentum symmetry down to $\mathds{Z}_4$ cannot occur; rather the spiral selection is continuous as a function of temperature. Even for a perfectly rotation invariant potential $\mathcal{H}_p$ (i.e. in the limit $\delta\rightarrow0$) a finite temperature transition associated with spontaneous momentum symmetry breaking $U(1)\rightarrow \mathds{Z}_4$ would be suppressed due to Mermin-Wagner theorem. Overall, the dominant effects of $\mathcal{H}_p$ and the momentum pinning makes a description in terms of a rank-2 U(1) gauge theory inaccurate and, consequently, the four-fold pinch points in the correlator $\mathcal{C}_{EE}({\bm k})$ fade drastically, see Fig.~\ref{fig:mc_results2}(b).

\begin{figure}[t]
\centering
\includegraphics[width=0.99\linewidth]{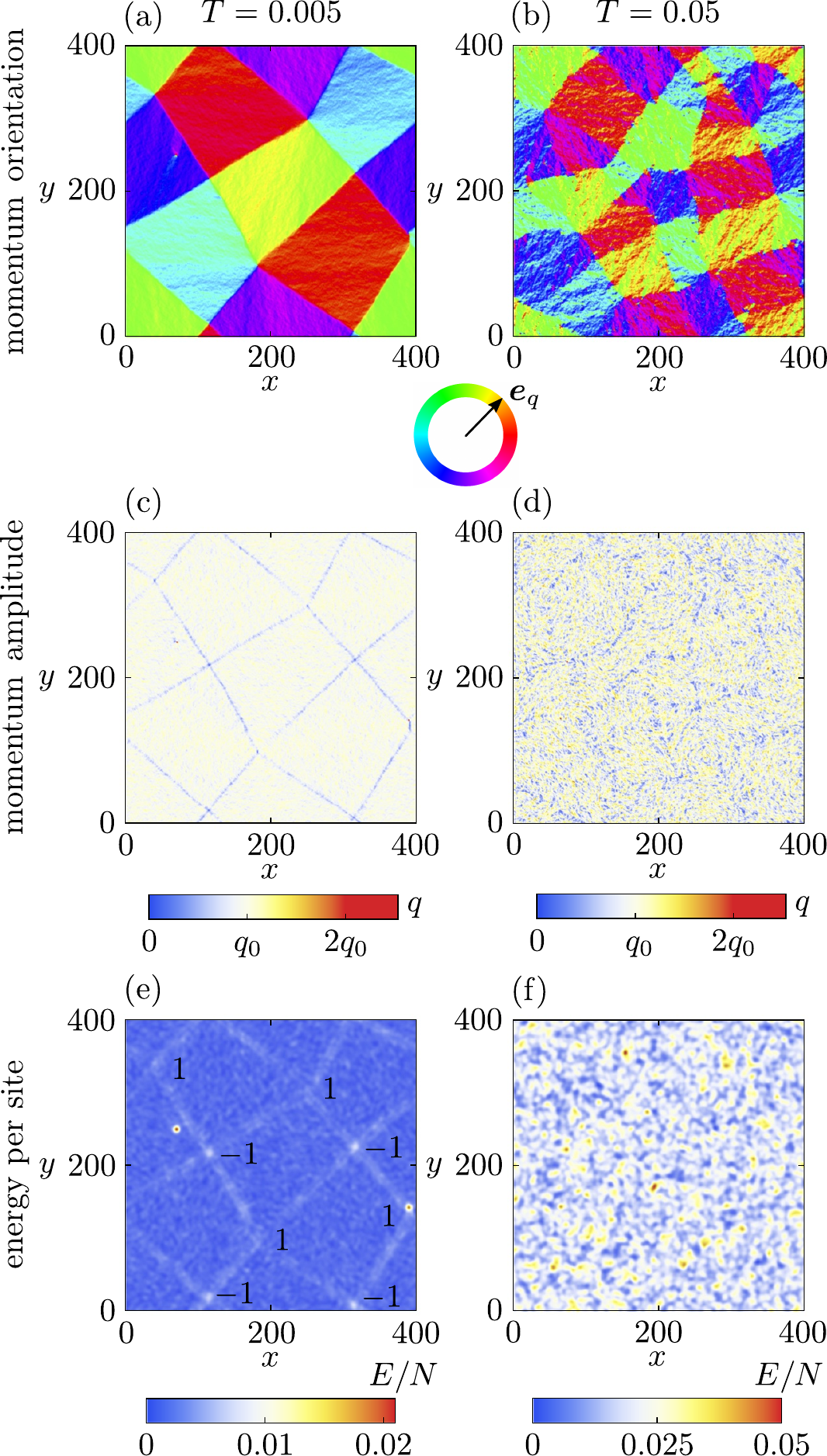}
\caption{Further details of the spin configurations from Fig.~\ref{fig1}(d) and Fig.~\ref{fig1}(e) at $T=0.005$ (left column) and $T=0.05$ (right column) . (a), (b) show the momentum orientation ${\bm e}_q={\bm q}/|{\bm q}|$, (c), (d) show the momentum amplitude $q=|{\bm q}|$ and (e), (f) show the local energies per site, smoothed by a Gaussian broadening with $\sigma=3$ lattice spacings. In (e) the momentum vortices and anitvortices are marked by their winding numbers $n=1$ and $n=-1$, respectively.}
\label{fig:momentum_mc}
\end{figure}

To discuss the system's behavior below the spiral transition at $T=0.08$ in more detail, we show in Fig.~\ref{fig:momentum_mc} the real-space momentum and energy distributions at $T=0.005$ and $T=0.05$. The formation of spiral patches and rigid networks of domain walls is clearly visible when comparing the local momentum directions in Fig.~\ref{fig:momentum_mc}(a) and Fig.~\ref{fig:momentum_mc}(b). Furthermore, since domain walls have reduced momenta, the network can be made visible by plotting the momentum amplitudes, see Fig.~\ref{fig:momentum_mc}(c) and Fig.~\ref{fig:momentum_mc}(d). While at $T=0.05$ only faint indications of domain walls are visible, they become very pronounced at $T=0.005$. The domain walls also show up as lines of enhanced energy, see Fig.~\ref{fig:momentum_mc}(e), however, in a less pronounced way as in the momentum amplitude. [In order to make the network of domains walls visible in the energy a Gaussian smoothing of the data with a standard deviation of $\sigma=3$ lattice spacings has been performed in Fig.~\ref{fig:momentum_mc}(e) and \ref{fig:momentum_mc}(f).] The spin configuration in Fig.~\ref{fig:momentum_mc}(f) also reveals that momentum antivortices ($n=-1$) cost more energy than momentum vortices ($n=1$). This is because $n=1$ vortices can reduce their energy by realizing spin patterns with an approximate circular symmetry near the vortex core. As an additional consequence of this freedom, we observe that $n=1$ vortices typically show larger deviations from $\pi/2$ angles between their domain walls than $n=-1$ vortices. Clearly, however, the excitations with the largest energy in Fig.~\ref{fig:momentum_mc}(f) are spin vortices which appear as narrow and high peaks in the energy landscape. This shows that topological defects in spin and momentum space occur on two different energy scales.

It should be emphasized, however, that at the lowest simulated temperatures ($T\approx0.005$) the typical time scales of thermal relaxation become exceedingly large. This is because of the rigidity of domain walls which cannot undergo any local moves but can only be modified when changing the spin state over the entire lattice. Therefore, it is possible (or even likely) that our numerically obtained low-temperature spin configurations do not represent thermal equilibrium but are rather metastable states. 

In summary, our low-temperature Monte-Carlo results reveal a well-defined network of narrow domain walls with a tendency for rectangular patterns minimizing the energy of momentum antivortices. While the domain sizes grow with decreasing temperature a transition into a phase where $U(1)$ momentum symmetry is spontaneously broken down to $\mathds{Z}_4$ (or lower symmetries) is not observed. In contrast to fluctuations in momentum space which take place on relatively small energy scales, spin vortices appear as massive and locally confined excitations.

\section{Discussion}
In this work, we have investigated the low-energy behavior of spiral spin liquids and have identified various unexpected properties and connections to other fields in physics. The key reason for the non-trivial behavior of spiral spin liquids lies in its effective U(1) degree of freedom which corresponds to the direction of the spiral momentum ${\bm q}$ on the degenerate ring-like ground state manifold. Unlike the elementary spin degree of freedom, the spiral momentum, which we define as a vector field ${\bm q}({\bm r})$ on a coarse-grained lattice, is subject to a curl-free condition, ${\bm \nabla}\times{\bm q}({\bm r})=0$. This has drastic consequences on the nature of excited spin configurations, particularly, we prove that momentum vortices can only have winding numbers equal or smaller than one while higher-winding-number vortex types are strictly forbidden.

Interestingly, even though the momentum field ${\bm q}(\bm r)$ is directly related to the spin degree of freedom via ${\bm q}(\bm r)={\bm \nabla}\Phi({\bm r})$ (where $\Phi$ is the spins' inplane angle) spin vortices and momentum vortices are independent excitations where the latter ones cost a much lower energy than the former. There is, hence, a temperature regime where thermal fluctuations mostly affect the momentum direction of spin spirals while the momentum amplitude is fixed near the ground state ring, leading to spin configurations with large densities of momentum vortices, see Fig.~\ref{fig1}(e). This is the spiral spin liquid regime whose existence we have numerically demonstrated for the classical square lattice $J_1$-$J_2$-$J_3$ XY model.

We have further demonstrated that the precise mechanism behind the momentum fluctuations in the spiral spin liquid bears striking similarities with elasticity theory of crystals, particularly, our low-energy continuum spin model can be directly mapped onto the shear modulus term in elasticity. However, due to the constrained nature of the momentum field, the topological defects of crystals -- disclinations and dislocations -- do not have any analogues in a classical spiral spin liquid. Rather, a momentum vortex with positive winding can be considered either as a bound pair of dislocations or a quadrupole of disclinations. Given the mapping between elasticity theory and rank-2 U(1) gauge theory for fractons one may alternatively describe the low-energy behavior of spiral spin liquids by a tensor gauge theory electrostatics subject to a generalized Gauss law. However, since deconfined charges (i.e. fractons) and dipoles of charges are strictly absent in spiral spin liquids, the thermal fluctuations of momentum vortices follow the low-energy behavior of an electrostatics tensor gauge theory where only fracton quadrupoles (or higher multipoles) are present. We make this connection explicit by numerically resolving the four-fold pinch points in the electric field correlator which are characteristic for these fracton theories.

Cooling down the spiral spin liquid, the fluctuations in the momentum amplitude continuously freeze out. This freezing is again controlled by the curl-free condition for ${\bm q}$ which dictates that the low-energy momentum antivortices must be formed by the intersections of straight and narrow domain walls separating regions of different momenta ${\bm q}$. We developed a classification scheme for momentum vortices based on the number of domain walls radiating from the vortex core and found that vortices with winding $n=-1$ have the smallest excitation energy when four domain walls with $\pi/2$ angles between them emanate from the center. Even though momentum vortices with $n=1$ can in principle be constructed in a smooth way without any domain walls they are incompatible with $n=-1$ vortices when arranging both in the same system. As a result, our low-temperature numerical simulations show rigid networks of domain walls connecting momentum vortices and antivortices where angles of $\pi/2$ between momenta in adjacent domains are energetically preferred. Interestingly, in such a network state no local excitations can be made in a way that domain walls are only modified, inserted or deleted in a finite lattice region. The simplest low-energy process of changing the domain wall configuration consists of shifting a domain wall parallel to itself over its full length. As a result of this non-local dynamics, equilibration times are increasing significantly as the temperature is decreased such that the system gets easily trapped in metastable states. A typical observation in our numerical outputs is that the domain wall network becomes more wide-meshed with decreasing temperature.

Our work sheds light on the nature of 2D spiral spin liquids, 
and also opens gateways to analyzing a plethora of related problems.
First, its generalization to 3D is highly non-trivial.
The degenerate spiral ground state wave-vectors in 3D can form a 1D ring, a 2D sphere, a sphere with punctures \cite{Gao2016NatPhys,Bergman2007NatPhys,Iqbal2018PRB}, or other manifolds with or without boundaries \cite{Yao2021Frontier}.
In each of these cases, the topological defects of momentum vectors are different,
and their classification will be an indispensable step in understanding 3D spiral spin liquids. 
Second, insights from this work may help us to understand the quantum version of spiral spin liquids. 
By identifying the classical spiral spin liquid as the electrostatics of a rank-2 U(1) theory, 
it is reasonable to speculate that the quantum model will at least carry some features of these effective theories.
We note that our construction seems to have a natural extension to the quantum model very similar to the higher-rank deconfined quantum criticality studied by Ma and Pretko \cite{Ma2018PhysRevB}.

We also highlight that the local momentum vortices are a new type of topological defect with a rather exotic curl-free constraint. Their detailed properties await more in-depth study.
For example, while spin vortices are only topological for XY spins, 
local momentum vortices are also well defined for Heisenberg spins. 
They have already been observed in studies of Heisenberg spiral spin liquid models on the honeycomb lattice by Shimokawa and Kawamura \cite{Shimokawa2019PhysRevLett}.
Closely related to this, the precise melting process of the rigid vortex network as temperature increases
represents an interesting statistical physics problem.

Our theoretical study has direct applications in experiment. 
Among various spiral spin liquid materials 
\MSC\ \cite{Gao2016NatPhys,Bergman2007NatPhys,Iqbal2018PRB},
${\mathrm{MgCr}}_{2}{\mathrm{O}}_{4}$ \cite{Bai2019PRL},
${\mathrm{CoAl}}_{2}{\mathrm{O}}_{4}$
 \cite{Zaharko2011PhysRevB},
 ${\mathrm{NiRh}}_{2}{\mathrm{O}}_{4}$
 \cite{Chamorro2018PhysRevMaterials},
 and 
 ${\mathrm{Ca}}_{10}{\mathrm{Cr}}_{7}{\mathrm{O}}_{28} $ \cite{Balz2016NatPhys,Balz2017PhysRevB,Sonnenschein2019PhysRevB,Kshetrimayum2020AOP,Pohle2021PhysRevB,pohle2017arxiv},
 our theory is most relevant to the 2D quantum spin liquid material 
  ${\mathrm{Ca}}_{10}{\mathrm{Cr}}_{7}{\mathrm{O}}_{28}$. 
It has been shown that in classical limit, this bilayer breathing kagome magnet can be mapped to a honeycomb lattice model at low temperatures, 
and exchange parameters place it very close to the spiral spin liquid phase on the honeycomb model \cite{Pohle2021PhysRevB,pohle2017arxiv}.
This phase is essentially the same as the one studied in our work. 
The spiral ring in the spin structure factor [Fig.~\ref{fig:mc_results1}(a)] has already been observed in neutron scattering \cite{Balz2016NatPhys,Balz2017PhysRevB}.
More direct experimental tests of our theory would be the search for four-fold pinch points in the correlator defined in Eq.~\eqref{EQN_EE_correlator} [Fig.~\ref{fig:mc_results1}(c)],
or taking direct snapshot of spin configurations using cutting-edge technology like electron holography \cite{dhar2021arXiv}. Despite the challenges of such studies, we strongly believe that the investigation of spiral spin liquids in real materials represents a fruitful research direction.

\section{Acknowledgement}
We thank Tokuro Shimokawa, Rico Pohle, Roderich Moessner, 
 Nic Shannon, Andriy Nevidomskyy for helpful discussions.
H.Y.   acknowledges the support of the National Science Foundation Division of Materials Research under the Award DMR-1917511.
Research at Rice University was also supported by the Robert A. Welch Foundation Grant No.~C-1818.   J. R.  acknowledges financial support from the German Research Foundation within the TRR 183 (project A04).

\bibliography{reference}

\end{document}